\documentclass[manuscript,screen]{acmart}

\AtBeginDocument{%
  \providecommand\BibTeX{{%
    \normalfont B\kern-0.5em{\scshape i\kern-0.25em b}\kern-0.8em\TeX}}}

\setcopyright{acmcopyright}
\copyrightyear{2025}
\acmYear{2025}
\acmDOI{XXXXXXX.XXXXXXX}

\usepackage[english]{babel}

\usepackage{amsmath}
\usepackage{graphicx}
\usepackage{spverbatim}
\usepackage{multirow}
\usepackage{listings}
\usepackage{booktabs}
\usepackage{tabularx}
\usepackage{longtable}
\newsavebox{\largestimage}

\usepackage{type1cm} 
\usepackage{xspace} 
\usepackage{balance} 
\usepackage[font={bf}, tableposition=top]{caption} 
\usepackage{subcaption} 
\usepackage{bold-extra} 
\usepackage[vlined,linesnumbered,ruled,noend]{algorithm2e} 
\usepackage{microtype} 
\usepackage{siunitx} 
\usepackage{xfrac} 
\usepackage{mathtools} 
\PassOptionsToPackage{hyphens}{url} 
\PassOptionsToPackage{bookmarks, pdftex, colorlinks=true, pagebackref=true, backref=page}{hyperref} 
\PassOptionsToPackage{square,numbers}{natbib} 
\usepackage{cleveref} 
\usepackage[hyperpageref]{backref} 
\usepackage{hyphenat} 
\usepackage[framemethod=TikZ]{mdframed}
\usepackage{enumitem} 

\newcolumntype{L}[1]{>{\raggedright\arraybackslash}p{#1}}

\usepackage[show]{chato-notes}

\newcommand{\spara}[1]{\smallskip\noindent\textbf{#1}}

\renewcommand*\backref[1]{\ifx#1\relax \else (Cited on #1) \fi}

\graphicspath{{images/}}

\DeclareUnicodeCharacter{2212}{-}

\newcommand{\rev}[1]{\textcolor{black}{#1}}

\begin{document}

\title{Measuring \rev{Online} Behavior Change with Observational Studies: a Review}

\author{Arianna Pera}
\email{arpe@itu.dk}
\affiliation{%
  \institution{IT University of Copenhagen}
  \streetaddress{Rued Langgaards Vej 7}
  \city{Copenhagen}
  \country{Denmark}
  \postcode{2300}
}

\author{Gianmarco De Francisci Morales}
\email{gdfm@acm.org}
\affiliation{%
  \institution{CENTAI}
  \streetaddress{Corso Inghilterra 3}
  \city{Torino}
  \country{Italy}
  \postcode{10138}}

\author{Luca Maria Aiello}
\email{luai@itu.dk}
\affiliation{%
  \institution{IT University of Copenhagen}
  \streetaddress{Rued Langgaards Vej 7}
  \city{Copenhagen}
  \country{Denmark}
  \postcode{2300}
}

\renewcommand{\shortauthors}{Pera et al.}

\begin{abstract}
Exploring \rev{online} \rev{behavior} change is imperative for societal progress in the context of 21st-century challenges.
We analyze 148 articles (2000-2023) \rev{focusing on behavior change in the digital space} and build a map that categorizes \rev{behaviors, behavior change detection methodologies, platforms of reference, and theoretical frameworks that characterize the analysis of online behavior change}.
Our findings reveal a focus on sentiment shifts, an emphasis on API-restricted platforms, and limited integration of theory.
We call for methodologies able to capture a wider range of \rev{behavior} types, diverse data sources, and stronger theory-practice alignment in the study of online \rev{behavior and its change}.
\end{abstract}



\maketitle

\section{Introduction}\label{sec:intro}

The ability to modify our behavior is crucial for enhancing our quality of life and advancing our societies, thus enabling us to adapt to the evolving world around us.
The study of behavior change in humans has been at the center stage of social psychology since the inception of \emph{behaviorism}, an experimental branch of the natural sciences research dedicated to the systematic observation of human behavior and the exploration of the drivers of its change~\cite{watson_psychology_1913}.
For nearly a century, behavioral psychologists have primarily evaluated \rev{behavior} change through inventories and surveys linked to controlled experiments~\cite{festinger1953research,staddon2014new,baum2017understanding} designed to investigate multiple types of behavior such as health habits~\cite{strecher_role_1986,elbel_promotion_2013}, civic participation~\cite{slemrod_taxpayer_2001}, and altruism~\cite{andreoni2010altruism}.

The scope of research concerning human \rev{behavior and its change} has significantly broadened with the widespread adoption of social media, which has led to a cultural shift where publicly sharing thoughts and experiences with online audiences has become commonplace~\cite{abramova2017understanding}.
The increasing availability of self-reported data from online platforms has opened new avenues for exploring fine-grained digital traces of behavior change~\cite{lazer2020computational}. \rev{This explorations are particularly flourishing in the realm of Computational Social Science, which is the main field of focus for our review}.
Although most studies on online behavior change are purely observational rather than experimental~\cite{zhang2020data} (with a few notable exceptions~\cite{kramer2014experimental}), they offer distinct advantages over experimental designs typical of the traditional social sciences.
First, social media users generate data organically within their daily experiences, thus mitigating the issues of ecological validity that often affect the artificial design of laboratory studies~\cite{cicourel1982interviews}.
Second, online data is rich in implicit signals left by users as they interact with the platform and with each other, which enable a comparison of overtly stated behavior with tacitly revealed preferences and opinions~\cite{bond2015quantifying}. 
Last, \rev{such studies} hold direct relevance to societal development because the Web not only documents behavior change but also catalyzes it, serving as a fertile ground for political discourse, online deliberation, and collective action on a wide range of issues~\cite{segerberg2011social,gerbaudo2012tweets,margetts2015political}.
\rev{These strengths, however, come with limitations. Observational studies afford little control over confounding variables, making causal inference necessarily more tentative than in controlled experiments. Consequently, insights derived from such studies should be interpreted with caution.}

\rev{Identifying behavior shifts} from online media is a challenging task due to the diverse range of practices people use online, and to the complex and partial ways in which these shifts can surface from digital footprints.
Furthermore, since the \rev{first developments} of behaviorism, there has been a proliferation of theoretical frameworks on behavior change~\cite{hull_principles_1943,rogers_diffusion_2014,prochaska_transtheoretical_2020} which contributes to shaping a nuanced and multi-faceted scenario.
Such complexity has led to a wealth of work that approaches the investigation of \rev{online behavior and its} change from multiple angles both methodologically and theoretically. Nevertheless, a comprehensive characterization of this research domain remains absent. As a result, three critical questions are worth considering:
\begin{enumerate}[leftmargin=*,label=\textbf{Q\arabic*.}]
    \item What types of \rev{behavior and behavior change} have been explored in online observational studies, and on which platforms?
    \item What techniques are most frequently used to quantify \rev{behavior and behavior change} from digital data?
    \item What are the theoretical frameworks on behavior change that have been used in support of online empirical studies?
\end{enumerate}

To answer these questions, this literature review aims to consolidate the disparate research concerning the quantification of \rev{behavior and its change} in individuals and groups from online data.
Our primary goal is to categorize such \rev{concepts}, and the methodologies and theories employed in their evaluation, thereby identifying existing gaps in the field and highlighting opportunities for future research.
\rev{In surveying the literature, we adopt a broad and multidisciplinary definition of behavior that includes not only observable actions but also cognitive and affective expressions such as attitudes, emotions, and opinions—key components of online interactions. 
While this extends beyond the strict scope of radical behaviorism~\cite{skinner1965science}, it resonates with perspectives from cognitive psychology~\cite{chomksy1959review} that emphasize internal mental states. 
Some theoretical extensions within behavior analysis, such as Relational Frame Theory~\cite{hayes2001relational}, have sought to expand the behavioral account to encompass complex verbal and cognitive phenomena. 
While we do not aim to reconcile these distinct theoretical traditions, we explicitly acknowledge their epistemological differences and engage with them critically.
Our selective integration reflects a pluralistic stance aligned with the multidisciplinary nature of Computational Social Science, where digital traces frequently reflect complex psychological phenomena.
Interpreting such data meaningfully often requires drawing on both behavioral and cognitive perspectives, without collapsing them into a single unified framework.
Moreover,} such a broader interpretation is particularly suited to social media platforms, which are designed to facilitate the sharing of internal thought processes~\cite{bazarova2015social}.

\rev{While closely related, the concepts of behavior and behavior change are not interchangeable.
In this review, we adopt a broad definition of behavior that includes both overt (observable) actions and covert (inferred) expressions such as attitudes, emotions, or opinions, as captured through digital traces.
Behavior change, in turn, refers to a temporal shift in these behaviors, whether at the individual or collective level.
This distinction is both methodological and theoretical: how behavior is measured, interpreted, and linked to broader outcomes can differ substantially from those used to assess its change, particularly in observational studies based on online data.}
To keep a focused and manageable scope, this review is limited to \emph{observational} studies, thereby excluding intervention-based experiments.
Arguably, observational studies constitute the majority of Computational Social Science research that focuses on digital media.

Our review is timely for two main reasons.
The first concerns the growing societal importance of online-mediated behavior change.
A growing consensus has developed within the scientific community that addressing some of the most defining challenges of the 21st century, such as climate change~\cite{watts20212020}, mass migrations~\cite{berchin2017climate}, and health crises~\cite{ronnerstrand2015trust}, necessitates grassroots, large-scale behavior change~\cite{van2009averting,otto2020social}.
\rev{Planetary-scale online media could be used to expedite collective behavioral transitions; however, this potential raises important ethical considerations related to surveillance, data privacy, platform regulation, and the protection of human rights.
Responsible research in this area must carefully balance the benefits of enabling behavior change with the risks of misuse, ensuring transparency and respect for individual autonomy.}
The latter is about an apparently narrowing window of opportunity for this stream of scientific exploration.
With major social media platforms limiting data access for research, it is crucial to evaluate whether such policy changes jeopardize behavior change research and to identify viable future alternatives.

\begin{figure}[h!]
  \centering
  \includegraphics[width=0.45\linewidth]{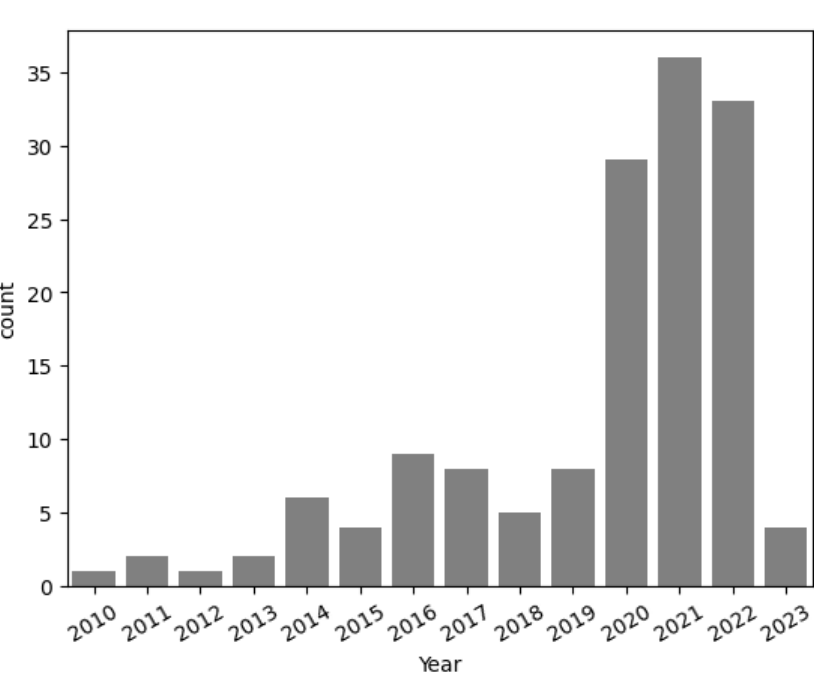}
  \caption{Number of retrieved papers by publishing \rev{year}.}
  \label{fig:years_numbers}
  \vspace{-10pt}
\end{figure}

We reviewed a total of 148 relevant articles published between 2000 and 2023 (\Cref{fig:years_numbers}, no articles retrieved between 2000 and 2010), which we retrieved using a thorough literature survey that we detail in \S\ref{sec:methods}.
Our findings reveal a significant bias in the existing literature, with the majority of studies concentrating on sentiment shifts as the primary form of \rev{behavior change}.
Most studies rely on online platforms that have been subject to increasingly rigorous API restrictions.
Furthermore, the empirical studies we examine seldom incorporate theoretical frameworks \rev{of behavior change from social and psychological sciences}.
This situation calls for innovative solutions, including: \emph{i)} \rev{the creation of ethically governed, privacy-conscious infrastructures that provide researchers with access to behavioral data, subject to appropriate safeguards and approvals}, \emph{ii)} the development of more comprehensive methodologies capable of encapsulating a wider array of human behavior types, and \emph{iii)} a more systematic incorporation of pertinent \rev{socio-psychological} theory into computational behavior studies.

\section{Taxonomy of studies on online behavior change} \label{sub:data_extr}
\rev{Behavior change research from online data typically involves three key entities: the \emph{environment} in which the behavior is enacted, the \emph{event} that prompts behavioral variation, and the \emph{behavior} itself.
This structure is inspired by behaviorist frameworks---particularly Applied Behavior Analysis (ABA)~\cite{cooper2007applied}—which provide a clear taxonomy linking environment, triggering events, and observable behaviors. 
In our study, we take inspiration from this tripartite organizational structure but explicitly extend it in terms of the types of behaviors considered. 
While ABA focuses on externally observable motor actions, our framework incorporates a broader range of behaviors made visible through digital traces, including affective, cognitive, and discursive expressions such as sentiment, opinion, and belief. 
We treat these internal states not as latent constructs preceding behavior, but as behaviors in their own right, following a multidisciplinary approach that is essential given the inherently multidisciplinary nature of Computational Social Science---the primary focus of this review.
This expanded behavioral lens is particularly suited for the study of online environments, where affective and cognitive expressions are made overt and tractable through text, interaction patterns, and platform features.
The \emph{environment} is, in the context of this study, defined as the online platform itself, which affords and constrains different types of behavioral expression. 
The \emph{event} refers to a contextual trigger (e.g., a political decision or health crisis), and the \emph{behavior} encompasses both actions and expressions that respond to or reflect that event. 
Each of these three} elements is characterized by multiple dimensions that are pertinent to the majority of relevant studies.
We will outline these dimensions, \rev{distinguishing between \emph{conceptual dimensions}, referred to what is being studied, and \emph{methodological dimensions}, referred to the analytical aspects of the study}.
To synthesize existing research, we extract all items relevant to these dimensions from the selected papers, and thus develop a taxonomy of observational studies for measuring \rev{behavior and its} change in online contexts.
The primary objective of this taxonomy is to discern connections between measurement methods, types of behavior and \rev{behavior} change, and socio-psychological dimensions of change. 
The resulting map of this research landscape can be used to pinpoint potential gaps in current research and assist scholars in categorizing new behavior change measurement methodologies.

\begin{table}[H]
    \centering
    \caption{\rev{Description of the types of online platforms retrieved from existing studies. Examples among the top-3 most cited works for each type are reported.}}
    \label{tab:table_platform_type}
    \begin{tabularx}{\textwidth}{l L{0.1\textwidth} S L{0.25\textwidth} L{0.45\textwidth}}
            \toprule
            \textbf{Var.} & \textbf{Label} & \textbf{N. works} & \textbf{Definition} & \textbf{Example}\\
            \midrule
            \multirow{20}{*}{\rotatebox{90}{\textbf{\rev{Type of online platform}}}}&social&106& social media&Twitter posts \cite{thelwall_sentiment_2011}\\ \cmidrule(lr){2-5}
            &forum &18&internet forum/online discussion site&mental-health support and non-mental-health subreddits during the initial stages of COVID-19 \cite{low_natural_2020}\\ \cmidrule(lr){2-5}
            &blog&8& blog websites data & web blog documents from wordpress.com \cite{jiang_topic_2011}\\ \cmidrule(lr){2-5}
            &e-learning&7&online learning platform & posts and click-stream log data from Moodle instances \cite{dascalu_before_2021}\\\cmidrule(lr){2-5}
            &review&5& website or platform that collects users thoughts about products & online employees' reviews from Glassdoor \cite{ren_psychological_2022}\\\cmidrule(lr){2-5}
            &search&5& search engine & click logs of Google News users \cite{liu_personalized_2010}\\\cmidrule(lr){2-5}
            &web traffic&3& internet log files & six network traces captured at a Pakistani ISP \cite{khattak_look_2014}\\\cmidrule(lr){2-5}
            &app&1& service usage data from service app & trip history data and station information data from bike sharing systems in Washington D.C. and Philadelphia \cite{lam_detecting_2019} \\\cmidrule(lr){2-5}
            &email&1& email services data & network of students' email communications \cite{uddin_impact_2014}\\\cmidrule(lr){2-5}
            &project \newline development&1&comments and change logs in software development & development artifacts for a specific release of Jazz \cite{licorish_understanding_2014}\\
            \bottomrule
    \end{tabularx}
\end{table}

\begin{enumerate}[leftmargin=*]
    \item \textbf{Environment}. \rev{ABA defines the environment of reference for a behavior as the full set of physical circumstances in which the organism exists. 
    Given that the focus of our review is on online behavior and behavior change, we focus on digital environments.} 
    \rev{In general,} the social context in which behavior is expressed plays a crucial role in defining its change, as outlined by the Social Cognitive Theory~\cite{bandura_social_1986}.
    Interactions in digital environments represent an evolution of such a concept. 
    In the context of our review, \rev{we consider online platforms as the digital environments in which behavior and behavior change manifest.}
    \begin{enumerate}
        \item \rev{\emph{Conceptual dimensions}}:
        \begin{enumerate}
            \item \rev{\emph{Type of online platform.} Category of online platform. \Cref{tab:table_platform_type} describes the volume distribution of such a dimension in the retrieved works.}
            \item \rev{\emph{Community.}} Online platforms can have a large audience with heterogeneous interests and norms, \rev{often being organized in interest-based groups}.
            It is therefore \rev{of interest to identify whether specific groups are more prone to manifest specific behavior changes}~\cite{bandura_social_1986}.
            We identify and extract the description of communities, groups, sections, or pages that are described as characterized by specific behaviors or behavior changes in the retrieved studies. 
            \rev{The results of such a process are detailed in \S\ref{subsec:environment}.}
        \end{enumerate}
    \end{enumerate}
    \item \textbf{Event}. \rev{When studying a given behavior, a shift} is typically triggered by an event or dynamic conditions within the surrounding context, rather than occurring spontaneously.
    This is well illustrated by the Health Belief Model in the medical field, where a cue or trigger is key to promoting health behaviors~\cite{janz_health_1984}. \rev{Such a framework, situated in cognitive psychology, is here integrated with ABA given that we situate our taxonomy in a highly multidisciplinary setting, in which we evaluate a broad range of behaviors including affective, cognitive, and discursive expressions.}
    \begin{enumerate}
        \item \rev{\emph{Conceptual dimensions}}:
        \begin{enumerate}
            \item \rev{\emph{Type of event.} The class of behavior change-triggering event.
            Volume distributions in relation to the retrieved works are summarized in \Cref{tab:table_event_type}.}
            \item \emph{Event source.} Whether the event is \emph{internal} or \emph{external} to the subject exhibiting the behavior.
            This distinction has been introduced by the Health Belief Model~\cite{janz_health_1984} and it is typical of intervention-based studies~\cite{fogg_behavior_2009}, but it also applies to observational studies.
            \rev{Detailed results are described in \S\ref{subsec_event}.}
        \end{enumerate}
        \item \rev{\emph{Methodological dimensions}}:
        \begin{enumerate}
            \item \emph{Event retrieval.} Whether the event is already known, given the available user-generated online data (e.g. post timestamps used to define a temporal phenomenon), or is latent and requires an indirect discovery via ad-hoc detection techniques.
            This dimension allows us to study the relationship between the complexity of events and their retrieval from data.
            \rev{Detailed results are described in \S\ref{subsec_event}.}
            \item \emph{Event detection technique.} The methods used to retrieve the event from the data. 
            Knowing how an event can be retrieved provides us with an overview of the most popular detection techniques applied for different event types and defines a set of guidelines to inform event retrieval in future behavior change studies. 
            \rev{Results are summarized in \Cref{tab:table_event_tech}}.
        \end{enumerate}
    \end{enumerate}
    
    \item \textbf{Behavior}. The behavior itself is the main focus of this kind of research.
    We adopt a broad definition of behavior that encompasses both the actions and thoughts that undergo some change.
    \rev{Such an inclusive definition fits well the Computational Social Science focus of the present review and} is commonly used in quantitative behavioral studies~\cite{heimlich_understanding_2008}\rev{, fitting well the characteristics of interactions on the digital space}.
    \begin{enumerate}
        \item \rev{Conceptual dimensions}:
        \begin{enumerate}
            \item \rev{\emph{Type of behavior}. Class of behavior analyzed in the retrieved studies.
            The volume distribution is reported in \Cref{tab:table_behavior_type}.}
            \item \emph{Behavior subject}: whether the behavior under study pertains to individuals or groups, as motivated by the dichotomous actor level described by \citet{heimlich_understanding_2008}.
            Group subjects can be additionally categorized either as collective entities or aggregations over individual behaviors. 
            \rev{Detailed results are reported in \S\ref{subsec:behavior}.}
            \item \emph{Behavior change subject.} Whether the behavior \emph{change} is measured at the individual or group level. 
            \rev{Detailed results are reported in \S\ref{subsec:behavior_change}.}
            \item \emph{Behavior object.} What the behavior is about; the object upon which the behavior acts or is defined. 
            The interest in behavior objects is partially grounded in the Object Relations Theory from psychoanalysis~\cite{kernberg_object_1995}, particularly in terms of how individuals form and maintain relationships with objects (including material objects, other people, but also abstract concepts) to shape behaviors and emotions. 
            \rev{Although this theoretical background differs epistemologically from the behaviorist foundations of the ABA model, which inspires our tripartite taxonomy, we draw on it as a valuable source for identifying behavioral dimensions, reflecting our multidisciplinary approach.}
            \rev{Results in terms of volume distribution in the retrieved works are reported in \Cref{tab:table_behavior_obj}.}
            \item \emph{Online/offline nature of behavior.} Whether the behavior is confined entirely within the online realm or it takes place offline and gets represented by online data.
            \rev{Detailed results are reported in \S\ref{subsec:behavior}.}
            \item \rev{\emph{Theoretical frameworks of behavior change.}} Psychological or social theories of behavior change relevant to the study, if explicitly mentioned by the authors. 
            \rev{Results are reported in \S\ref{subsec:theories_behavior_change}.}
        \end{enumerate}
        \item \rev{Methodological dimensions}:
        \begin{enumerate}
            \item \emph{Behavior proxy measure.} The methods used to quantify the behavior of interest. 
            \rev{The volume distribution of retrieved works is reported in \Cref{tab:table_behavior_proxy}.}
            \item \emph{Behavior change detection technique.} The methods used to identify and quantify a \emph{change} in behavior. 
            \rev{The volume distribution of retrieved works is reported in \Cref{tab:table_behavior_change_tech}.}
            \item \emph{Temporal span of behavior change.} 
            The temporal characteristics of the behavior change, which can be discrete or gradual, as described by~\citet{heimlich_understanding_2008}. 
            In particular, change is measured in discrete terms when comparing two consecutive time frames or by identifying change points in a temporal signal. 
            It is measured over a continuous spectrum when more fine-grained temporal measurements are performed.
            \rev{Detailed results are reported in \S\ref{subsec:behavior_change}.}
        \end{enumerate}
    \end{enumerate}
\end{enumerate}
\noindent After reviewing all the relevant papers, we populate the taxonomy and group studies according to the values that the different dimensions assume empirically.
In \S\ref{sec:results}, we provide a detailed account of all the reviewed work across the entities and dimensions considered.

\section{Map of methods for the quantification of online behavior change}\label{sec:results}
We present a comprehensive map of the current research landscape concerning online behavior change measurement, based on the previously described taxonomy. Our analysis begins with an exploration of the online \textbf{environments} that have been considered in the literature, followed by a review of the \textbf{events} studied as potential catalysts for behavior change. We conclude with a description of the various perspectives \rev{offered by existing works in terms of} \textbf{behavior} and \textbf{behavior change}\rev{, together with references to the \textbf{theoretical frameworks of behavior change} referenced by the studies}. It is worth noting that the research in this area tends to favor certain popular configurations, resulting in some branches of our taxonomy being applicable to a large number of studies, too extensive to be fully detailed in the main text (for instance, studies using Twitter as the primary online environment). Consequently, a comprehensive mapping of \rev{all analyzed} studies is provided in the Supplementary Materials.

\begin{table}
    \centering
    \caption{\rev{Description of the types of behavior-change-triggering events retrieved from existing studies. Examples among the top-3 most cited works for each type are reported.}}
    \label{tab:table_event_type}
    \begin{tabularx}{\textwidth}{l L{0.1\textwidth} S L{0.35\textwidth} L{0.3\textwidth}}
            \toprule
            \textbf{Var.} & \textbf{Label} & \textbf{N. works} & \textbf{Definition} & \textbf{Example}\\
            \midrule
            \multirow{32}{*}{\rotatebox{90}{\textbf{\rev{Type of event}}}}&disease&52&The event is directly or indirectly related to the outbreak or to the evolution of a disease&evolution of COVID-19 pandemics \cite{valdez_social_2020}\\ \cmidrule(lr){2-5}
            &news&17&The event is either a natural disaster, a war, a shooting, a crisis, a fact related to the life of famous people&suicide and depression-related events \cite{mcclellan_using_2017}\\ \cmidrule(lr){2-5}
            &politics&7&The event is related to elections or political facts&2012 Alberta General Elections \cite{makazhanov_predicting_2013}\\ \cmidrule(lr){2-5}
            &policy&6&The event is about the introduction of policies from leaders (either politicians or people leading a given community)&Goods and Services Tax (GST) implementation by Indian government \cite{singh_smart_2020}\\ \cmidrule(lr){2-5}
            &mixed events&4&Multiple events, related to different categories, are taken into account&different events happening in the world related to policy, movie, accident, terrorism and sport \cite{naskar_emotion_2020}\\ \cmidrule(lr){2-5}
            &special days&4&The event is a holiday or a day with some special characteristics (e.g., sales events)&sales events in the UK \cite{ibrahim_decoding_2019}\\ \cmidrule(lr){2-5}
            &social campaign&3&The event is a content-diffusion developed on social media&\#metoo movement \cite{foster_metoo_2022}\\ \cmidrule(lr){2-5}
            &exposure to online content&3&The event is the exposure to specific online content (e.g., posts) which is supposed to trigger a change&removal of undesirable comments \cite{srinivasan_content_2019}\\\cmidrule(lr){2-5}
            &personal event&3&The event is a phenomenon affecting the subject of the behavior, either cognitively or physically&moments of cognitive change \cite{kushner_bursts_2020}\\\cmidrule(lr){2-5}
            &\rev{badge earning}&\rev{3}&\rev{The event is about obtaining a qualification badge on online platforms}&\rev{\textit{Good Article} status achievement on Wikipedia\cite{zhang_crowd_2017}}\\\cmidrule(lr){2-5}
            &\rev{sport}&\rev{2}&\rev{The event is a sport competition}&\rev{broadcasting of the Winter Olympic Games \cite{hennig_big_2016}}\\\cmidrule(lr){2-5}
            &\rev{interaction}&\rev{1}&\rev{The event is the support/feedback received online}&\rev{help received from Amazon customer support \cite{ahmed_prediction_2022}}\\\cmidrule(lr){2-5}
            &\rev{project development}&\rev{1}&\rev{The event is a product development process}&\rev{phases of developing a software project \cite{licorish_understanding_2014}}\\ 
            \bottomrule
    \end{tabularx}
\end{table}

\begin{table}
    \centering
    \caption{\rev{Description of the techniques used in existing studies to detect behavior change-triggering events. Examples among the top-3 most cited works for each type are reported.}}
    \label{tab:table_event_tech}
    \begin{tabularx}{\textwidth}{l L{0.1\textwidth} S L{0.3\textwidth} L{0.35\textwidth}}
            \toprule
            \textbf{Var.} & \textbf{Label} & \textbf{N. works} & \textbf{Definition} & \textbf{Example}\\
            \midrule
            \multirow{20}{*}{\rotatebox{90}{\textbf{\rev{Event detection technique}}}}&fixed dates of interest&82&The time of the event is known (common knowledge)&COVID-19 outbreak in 2020 (Jan-Apr) compared to the same period in 2019 \cite{low_natural_2020}\\\cmidrule(lr){2-5}
            &meta data&7&The event is embedded in the data&time-labeled nomination and promotion to Good Article in Wikipedia data \cite{zhang_crowd_2017}\\\cmidrule(lr){2-5}
            &time series anomaly detection&5&The event is defined as a peak in time series of data&deconvolution of news volume time series to detect events \cite{tsytsarau_dynamics_2014}\\\cmidrule(lr){2-5}
            &known hashtags&4&The event is retrieved from hashtags&hashtags such as \textit{Obama} and \textit{GOP} to identify political events \cite{amelkin_distance_2019}\\\cmidrule(lr){2-5}
            &manual annotations&4&The event is retrieved by manually annotating data&negative tones comments annotated via Amazon Mechanical Turk\cite{sridhar_estimating_2019}\\\cmidrule(lr){2-5}
            &regular expressions&2&The event is retrieved by observing the presence of regular expressions&moments of cognitive change expressed with comments such as ``I feel much better now'' \cite{kushner_bursts_2020}\\\cmidrule(lr){2-5}
            &known mentioned account&1&The event is defined by the activity of a known account&interaction with the customer help account \textit{AmazonHelp} on Twitter \cite{ahmed_prediction_2022}\\
            \bottomrule 
    \end{tabularx}
\end{table}

\begin{table}[ht]
    \centering
    \caption{\rev{Description of the types of behaviors analyzed by existing studies. Examples among the top-3 most cited works for each type are reported.}}
    \label{tab:table_behavior_type}
    \begin{tabularx}{\textwidth}{l L{0.1\textwidth} S L{0.35\textwidth} L{0.3\textwidth}}
            \toprule
            \textbf{Var.} & \textbf{Label} & \textbf{N. works} & \textbf{Definition} & \textbf{Example}\\
            \midrule
            \multirow{18}{*}{\rotatebox{90}{\textbf{\rev{Type of behavior}}}}&emotion / sentiment&64&The sentiment or the emotions of a subject&public sentiment towards specific targets \cite{tan_interpreting_2014}\\ \cmidrule(lr){2-5}
            &interest&56&Interest of users, both in terms of topics and numbers of contributions&consumption of news articles \cite{liu_personalized_2010}\\ \cmidrule(lr){2-5}
            &attitude&26&How users think or feel with respect to a topic/person/event, including stance&mental associations of words in relation to sinophobic behavior \cite{tahmasbi_go_2021}\\ \cmidrule(lr){2-5}
            &spatial&14&How users move around in real-life&mobility pattern of population affected by haze events \cite{kibanov_mining_2017}\\\cmidrule(lr){2-5}
            &community dynamics&9&How online communities are characterized in terms of interactions, participation and structure& interactions between students in an email communication network \cite{uddin_impact_2014}\\\cmidrule(lr){2-5}
            &mental state&6&Psychological situation, often negative, affecting users&suicide-related behavior emerging from social media posts \cite{vioules_detection_2018}\\\cmidrule(lr){2-5}
            &real-life habits&2&Real-life actions of users that are part of their routine&alcohol drinking stage of social media users \cite{liu_assessing_2017}\\
            \bottomrule
    \end{tabularx}
\end{table}

\begin{table}[ht]
    \centering
    \caption{\rev{Description of the types of objects of behaviors analyzed by existing studies. Examples among the top-3 most cited works for each type are reported.}}
    \label{tab:table_behavior_obj}
    \begin{tabularx}{\textwidth}{l L{0.1\textwidth} S L{0.35\textwidth} L{0.3\textwidth}}
            \toprule
            \textbf{Var.} & \textbf{Label} & \textbf{N. works} & \textbf{Definition} & \textbf{Example}\\
            \midrule
            \multirow{12}{*}{\rotatebox{90}{\textbf{\rev{Behavior object}}}}&event&78&The behavior is defined with respect to an event or related topics&sentiment with respect to emerging events \cite{thelwall_sentiment_2011}\\\cmidrule(lr){2-5}
            &online content&17&The behavior is defined with respect to online posts or information&interest with respect to e-learning content \cite{dascalu_before_2021}\\\cmidrule(lr){2-5}
            &people&16&The behavior is referred to its own subject or to other people&sentiment with respect to election candidates \cite{chaudhry_sentiment_2021}\\\cmidrule(lr){2-5}
            &product&1&The behavior is defined with respect to a commercialized object&sentiment with respect to products and services of online retail brands \cite{ibrahim_decoding_2019}\\\cmidrule(lr){2-5}
            &place&1&The behavior is defined with respect to a location&mobility patterns in relation to the presence of green spaces \cite{liu_categorization_2020} \\
            \bottomrule
    \end{tabularx}
\end{table}

\begin{table}[ht]
    \centering
    \caption{\rev{Description of the proxy measures used to extract the behaviors analyzed by existing studies. Examples among the top-3 most cited works for each type are reported.}}
    \label{tab:table_behavior_proxy}
    \begin{tabularx}{\textwidth}{l L{0.1\textwidth} S L{0.33\textwidth} L{0.33\textwidth}}
            \toprule
            \textbf{Var.} & \textbf{Label} & \textbf{N. works} & \textbf{Definition} & \textbf{Example}\\
            \midrule
            \multirow{26}{*}{\rotatebox{90}{\textbf{\rev{Behavior proxy measure}}}}&count-based&70&The behavior is measured by tracking online activity numerically&number of clicks on news articles \cite{liu_personalized_2010}\\\cmidrule(lr){2-5}
            &dictionary-based&64&The behavior is measured by taking into account the occurrences of words in a specific lexicon&VADER lexicon for sentiment analysis \cite{valdez_social_2020}\\\cmidrule(lr){2-5}
            &embedding-based&25&The behavior is measured by extracting word or sentence embeddings from text&GloVe and TF-IDF-based embeddings to detect stress \cite{li_modeling_2020}\\\cmidrule(lr){2-5}
            &classifier&17&The behavior is measured through a machine learning model that takes in input some basic features such as term frequencies&multiple ML algorithms such as the SVM, RF, decision tree (DT), CNN, and Tree-CNN to detect emotions \cite{rosa_event_2020}\\\cmidrule(lr){2-5}
            &topic-based&13&The behavior is measured via topic modeling&dynamic multinomial Dirichlet mixture user clustering topic model \cite{zhao_explainable_2016}\\\cmidrule(lr){2-5}
            &location-based&12&The behavior is related to motion and is measured by taking into account geotagged user-generated content&geo-tagged Instagram posts as location \cite{rodigruez_dominguez_sensing_2017}\\\cmidrule(lr){2-5}
            &network-based&11&The behavior is measured via structural characteristics of the social network&number of micro-clusters in a tweet graph around a given topic as a measure of topic diversity\cite{hashimoto_analyzing_2021}\\\cmidrule(lr){2-5}
            &manual annotation&1&The presence of the behavior is manually annotated&experts annotations of professional learning community discourse \cite{scherz_whatsapp_2022}\\
            \bottomrule
    \end{tabularx}
\end{table}

\begin{table}[ht]
    \centering
    \caption{\rev{Description of the techniques of detection for the behavior change analyzed by existing studies. Examples among the top-3 most cited works for each type are reported.}}
    \label{tab:table_behavior_change_tech}
    \begin{tabularx}{\textwidth}{l L{0.1\textwidth} S L{0.35\textwidth} L{0.3\textwidth}}
            \toprule
            \textbf{Var.} & \textbf{Label} & \textbf{N. works} & \textbf{Definition} & \textbf{Example}\\
            \midrule
            \multirow{20}{*}{\rotatebox{90}{\textbf{\rev{Behavior change detection technique}}}}&simple quantitative&55&The technique for behavior change detection involves quantitative measures that are compared or analyzed. No statistical assessment is involved&percentage of negative or positive tweets increasing for more than 50\% \cite{tan_interpreting_2014}\\\cmidrule(lr){2-5}
            &visual analysis&53&Visual analysis of behavior trends, visual topic analysis or word clouds&visual evolution of topic ranking over time \cite{valdez_social_2020}\\\cmidrule(lr){2-5}
            &statistical method&45&Any technique involving some statistical measures such as test for the significance of difference or of model coefficients&Wilcoxon signed ranks test comparing differences in sentiment strength before and after events \cite{thelwall_sentiment_2011}\\\cmidrule(lr){2-5}
            &learning&32&Any technique which involves the identification of values for some unknown parameters&linear smooth kernel regression of emotions \cite{iglesias-sanchez_contagion_2020}\\\cmidrule(lr){2-5}
            &time series&23&Techniques related to time series analysis, either in terms of structural properties, time series modeling or anomaly detection methods&deviation from the time series forecast of number of Tweets about a given topic with ARIMA \cite{mcclellan_using_2017}\\\cmidrule(lr){2-5}
            &network-based&6&Techniques based on the comparison of network states or on the use of models to analyze structural parameters&Stable Group Changes Identification to study group dynamics in blogs \cite{zygmunt_achieving_2020}\\
            \bottomrule
    \end{tabularx}
\end{table}

\subsection{Environment} \label{subsec:environment}
\subsubsection{Executive summary}

\begin{figure}[htbp]
  \centering
     \begin{subfigure}[b]{0.71\textwidth}
         \centering
         \includegraphics[width=\textwidth]{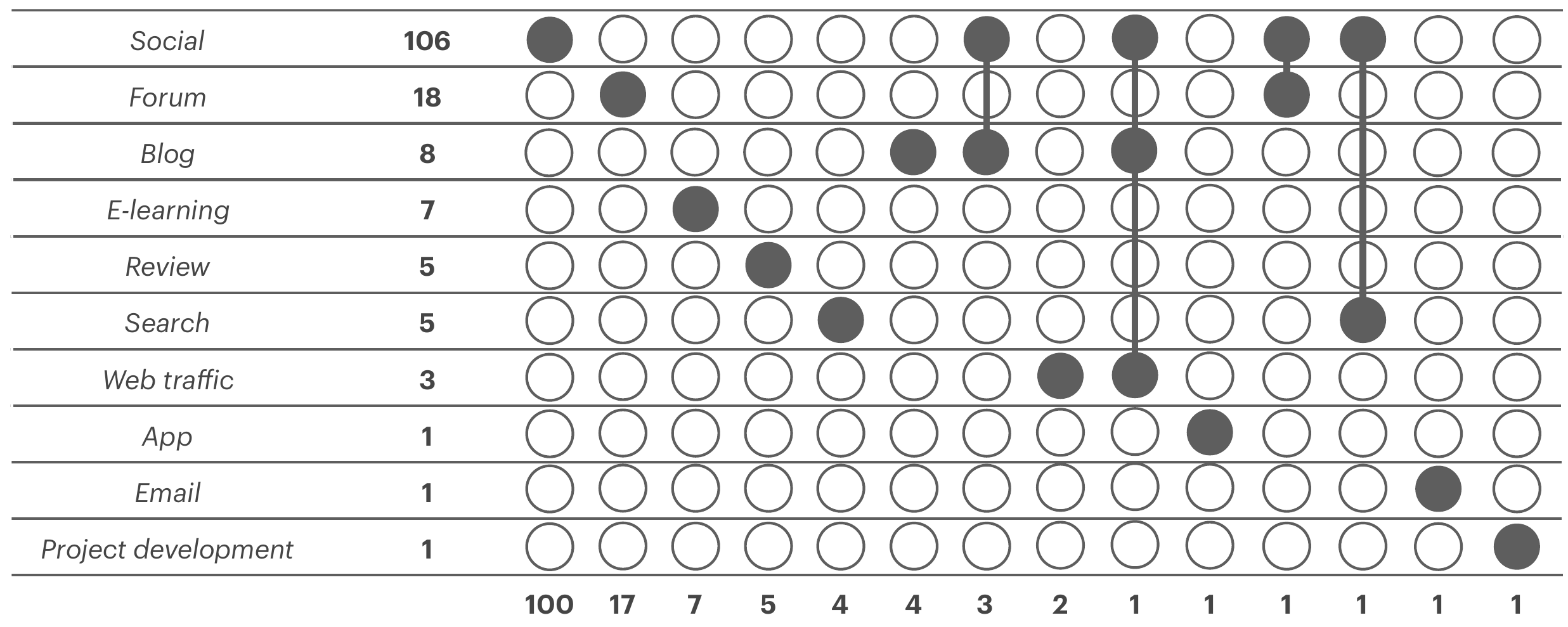}
         \caption{Online platforms.}
         \label{fig:platform_gen}
     \end{subfigure} \\[4pt]
     \begin{subfigure}[b]{0.58\textwidth}
         \centering
         \includegraphics[width=\textwidth]{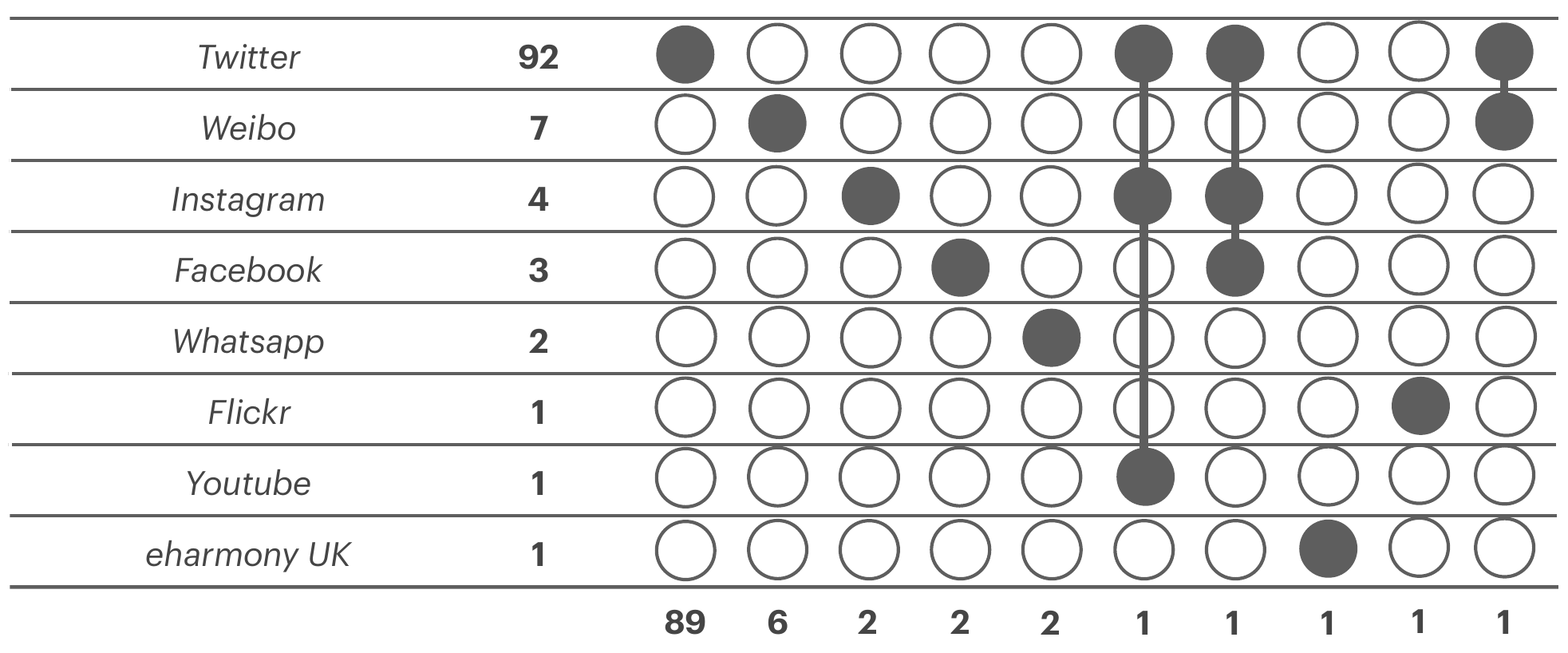}
         \caption{\rev{Focus on} Social media.}
         \label{fig:social_media}
     \end{subfigure} \\[4pt]
     \begin{subfigure}[b]{0.62\textwidth}
         \centering
         \includegraphics[width=\textwidth]{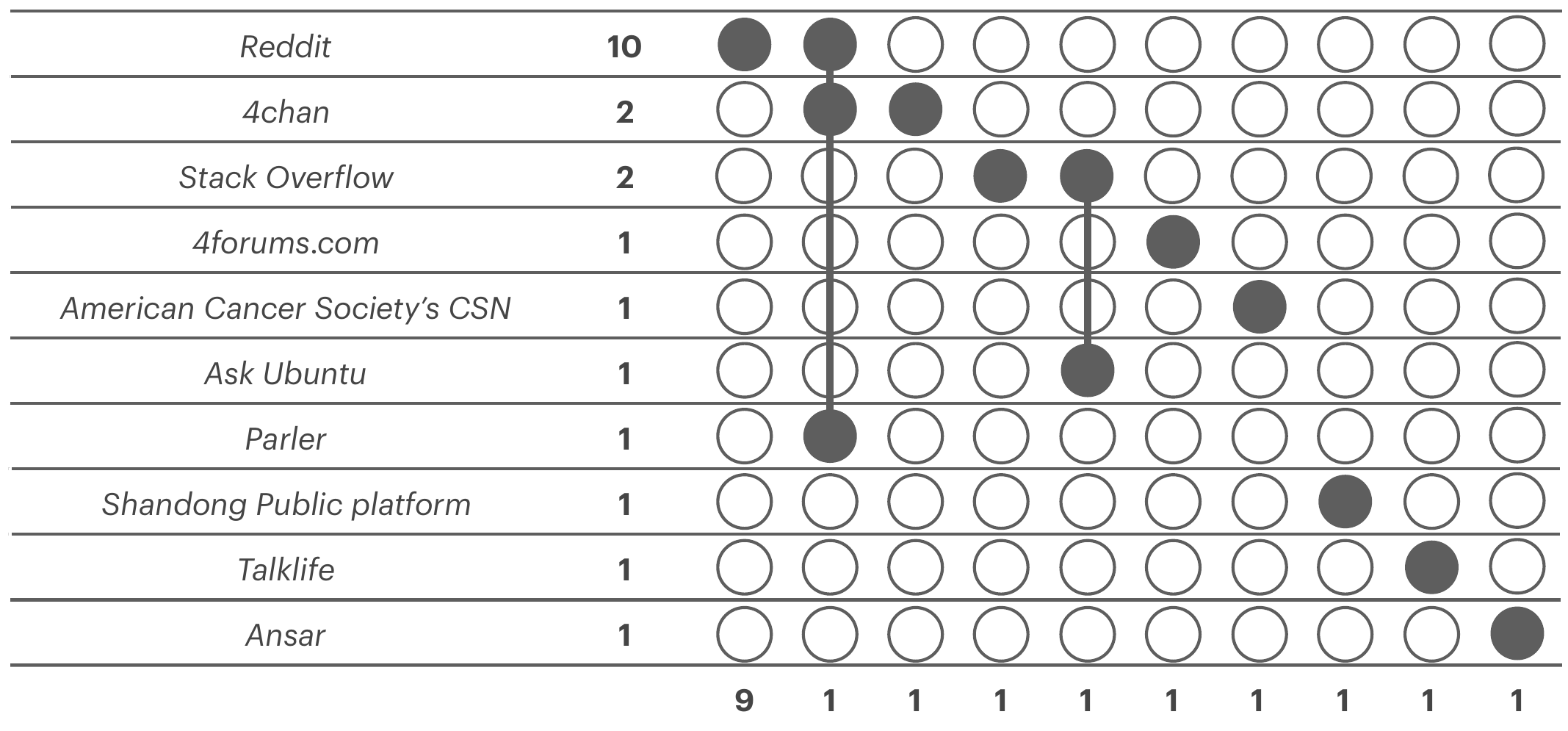}
         \caption{\rev{Focus on} Forums.}
         \label{fig:forum}
     \end{subfigure} 
        \caption{\rev{Number of works by type of online platform (a) and in-depth details on the specific social media (b) and forum platforms (c) analyzed. Dot connectors characterize the simultaneous presence of multiple characteristics.}}
        \label{fig:platforms}
        \vspace{-10pt}
\end{figure}

A significant proportion of research on the quantification of online \rev{behaviors and} behavior change focuses on social media platforms (\rev{cf. \Cref{tab:table_platform_type} and \Cref{fig:platform_gen}}).
Approximately 68\% of these studies exclusively use social media data, while a further 4\% use it in conjunction with other platforms such as blogs, web traffic, online forums, and search engines. Twitter is by far the predominant social media platform of choice, taking up 87\% of social media-based studies and 62\% of all studies (\Cref{fig:social_media}).
A considerable volume of research also leverages data from online forums, with these platforms contributing to at least a portion of the data in 12.2\% of studies.
Among them, Reddit is the most commonly analyzed forum, featuring in 55.6\% of forum-based research (\Cref{fig:forum}).

\subsubsection{Detailed mapping} 
Twitter is by far the most popular data source of reference.
This platform is frequently subjected to independent evaluation, and this examination is predominantly categorized into three primary focus areas in terms of data retrieval. 
The first considers a time span of interest, such as in \citet{tan_interpreting_2014} and in \citet{chaudhry_sentiment_2021}, the second investigates specific entities (accounts, such as in \citet{makazhanov_predicting_2013} and in \citet{jiang_modeling_2015}) or locations, such as in \citet{kibanov_mining_2017}) and the third extracts relevant data with a keyword-based strategy, such as in \citet{wang_anomaly_2015, mcclellan_using_2017} and \citet{crocamo_surveilling_2021} or a topic modeling-based strategy, such as in \citet{sidana_health_2018}.
There are a small number of examples of inter-platform investigations exploring Twitter in connection with other sources: a few studies consider a multi-platform social media analysis, namely by exploring Twitter together with Instagram and Youtube \cite{iglesias-sanchez_contagion_2020}, with Weibo \cite{rosa_event_2020} or with Instagram and Facebook \cite{jindal_intend_2018}. 
Some other works take into account inter-platform studies beyond the pure social media, by exploring the intersection of Twitter with Google search activities\cite{sharpe_evaluating_2016, evers_covid-19_2021, sajjadi_public_2021}, by investigating also chat platforms \cite{pellert_dashboard_2020, yu_temporal_2020}, by including analyses on blog portals \cite{denecke_how_2013, tsytsarau_dynamics_2014} and by expanding the focus to online forums \cite{tahmasbi_go_2021} or to mobility data \cite{mejova_youtubing_2021}.
There are also some contributions focused mainly towards different social media platforms, namely Weibo \cite{wang_spatiotemporal_2019, rosa_event_2020, liu_categorization_2020, hu_detecting_2020, cao_learning_2021, su_public_2021, han_weibo_2022}, Instagram \cite{rodigruez_dominguez_sensing_2017, jindal_intend_2018, redondo_hybrid_2020, iglesias-sanchez_contagion_2020}, Facebook \cite{hennig_big_2016, jindal_intend_2018, shepherd_domestic_2021}, Whatsapp \cite{chen_exploring_2022, scherz_whatsapp_2022}, Flickr \cite{jing_fine-grained_2020} and eharmony \cite{dinh_computational_2022}.  

The line of research based on online forums data is primarily focused on Reddit and, in particular, on event-specific subreddits such as \emph{politics} \cite{tachaiya_sentistance_2021}, \emph{FortniteBR} \cite{ryan_lock_2020}, \emph{OpiatesRecovery} \cite{petruzzellis_relation_2023, balsamo_pursuit_2023}), on mental-health related communities \cite{low_natural_2020}, on location-based communities \cite{basile_how_2021, liu_monitoring_2021}, on very popular general-topic communities such as \emph{AskReddit}, \emph{todayilearned} and \emph{science} \cite{almerekhi_are_2020, almerekhi_investigating_2022} and on the \emph{ChangeMyView} community \cite{srinivasan_content_2019}.
Other forum platforms that are investigated, even though to a lesser extent, are 4chan \cite{tahmasbi_go_2021, tachaiya_sentistance_2021}, Stack Overflow \cite{yanovsky_one_2021, hoernle_phantom_2022}, 4forums.com \cite{sridhar_estimating_2019}, Ask Ubuntu \cite{yanovsky_one_2021}, Parler \cite{tachaiya_sentistance_2021}, Talklife \cite{kushner_bursts_2020} and Ansar \cite{theodosiadou_change_2021}.
Some works focus on more niche and domain-specific sources: \citet{bui_temporal_2016} analyze the American Cancer Society's CSN, an online network for cancer patients, survivors, and caregivers, while \citet{sun_new_2021} explore the Shandong Environmental Public Prosecution platform, a Chinese public network platform to complain about environmental pollution in Shandong Province. 

A portion of the remaining research is based on e-learning platforms, by taking into account posts of enrolled users \cite{peng_topic_2020, tao_data_2022}, activity logs \cite{shi_temporal_2020, zhang_measuring_2021, zhao_academic_2021, dermy_dynamic_2022} or both \cite{dascalu_before_2021}.
Blogs are taken into account in terms of Wikipedia pages \cite{sharpe_evaluating_2016, zhang_crowd_2017} or in a more general form \cite{jiang_topic_2011, denecke_how_2013, tsytsarau_dynamics_2014, jindal_intend_2018, zygmunt_achieving_2020, lu_agenda-setting_2022}. 
Some analyses are also performed in terms of users' opinions on specific products or entities: \citet{kim_analysis_2021} and \citet{rabiu_drift_2023} consider Amazon's customer reviews as a data source, \citet{ren_psychological_2022} evaluate employees' reviews on Glassdoor, \citet{steinke_sentiment_2022} investigate IMDB movies reviews and \citet{karatas_novel_2022} study community evolution on Yelp.
Search engine data, especially in terms of Google \cite{liu_personalized_2010, mavragani_infoveillance_2018, evers_covid-19_2021, sajjadi_public_2021} and Naver or Daum \cite{jung_study_2021} searches, and web traffic logs \cite{khattak_look_2014, sharpe_evaluating_2016, epure_modeling_2017} are analyzed only to a minor extent.
Finally, app data \cite{lam_detecting_2019}, email networks \cite{uddin_impact_2014} and project development logs \cite{licorish_understanding_2014} represent very niche domains of investigation. 

\subsection{Events} \label{subsec_event}

\subsubsection{Executive summary}
\begin{figure}[htbp]
  \centering
     \begin{subfigure}[b]{0.69\textwidth}
         \centering
         \includegraphics[width=\textwidth]{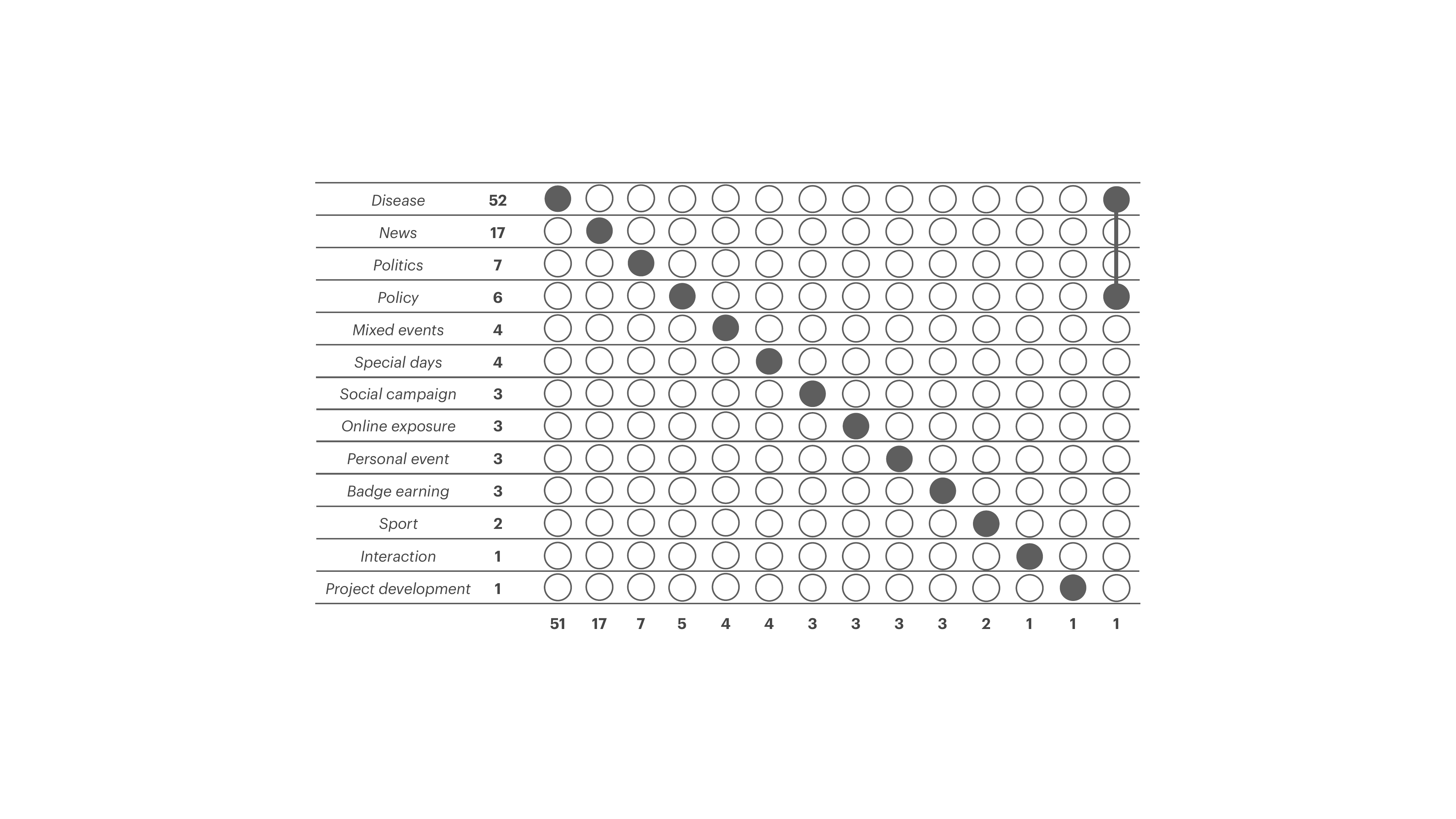}
         \caption{Event type.}
         \label{fig:event_type}
     \end{subfigure} \\[4pt]
     \begin{subfigure}[b]{0.58\textwidth}
         \centering
         \includegraphics[width=\textwidth]{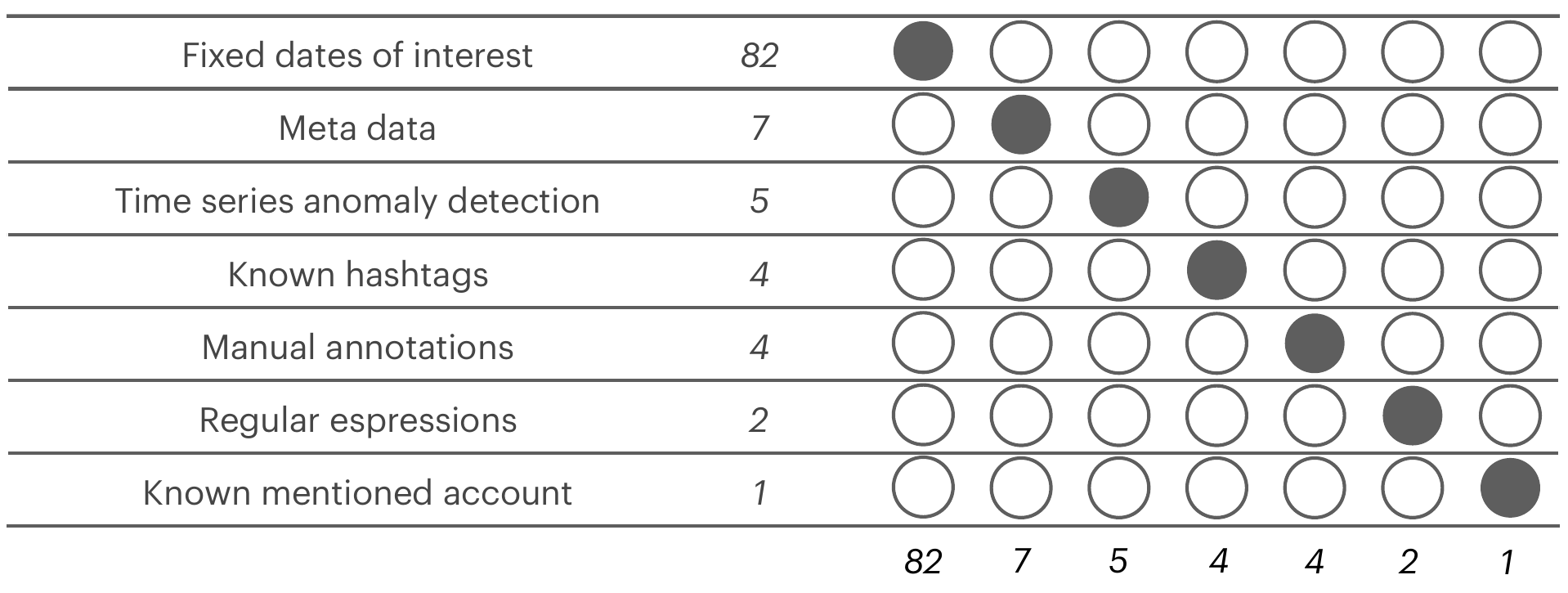}
         \caption{Event detection technique.}
         \label{fig:event_tech}
     \end{subfigure}   
  \caption{\rev{Number of works by type of triggering event (a) and technique used by scholars for event detection (b). Dot connectors characterize the simultaneous presence of multiple characteristics.}}
\end{figure}

We found that 70\% of behavior change \rev{operationalization} studies explicitly mention a potential triggering event \rev{for behavior change}.
The remaining 30\% of the research either focuses on broader topics around which behavior change is explored, or analyses available data streams not connected to any specific topic.
Half of the event-focused studies concentrate on disease-related events, with a particular emphasis on COVID-19, which accounts for 96\% of these disease-related cases (\rev{cf. \Cref{tab:table_event_type} and \Cref{fig:event_type}}).
Additionally, a substantial body of work investigates \rev{shifts} in relation to news events (16\%), political events (7\%), and policy introductions (6\%).
Notably, 97\% of studies that consider an event as a catalyst for behavior change examine events that are external relative to the behavior\rev{-adopting} subject.
Furthermore, 91\% of event-based studies derive the event directly from user-generated content (\rev{cf. \Cref{tab:table_event_tech} and \Cref{fig:event_tech}}), typically in connection with known phenomena of interest occurring at a specific time.

\subsubsection{Detailed mapping}
Given the impact the 2020 pandemic has had worldwide, it is understandable that a consistent amount of contributions starting from 2020 study COVID-19 as a behavior change-triggering event.
The majority of works consider the overall pandemic effect (as, for instance, \citet{valdez_social_2020, tahmasbi_go_2021, liu_monitoring_2021}) but some research narrows down the focus on specific sub-events, such as lockdowns \cite{sciandra_covid-19_2020, mejova_youtubing_2021, aljeri_big_2022}, vaccine introduction \cite{li_dynamic_2022}, governmental mandates \cite{yeung_face_2020, mejova_authority_2023}, and information diffusion around the topic \cite{evers_covid-19_2021, chen_exploring_2022}.
Other disease-related behavior changes are investigated with a focus on ebola \cite{ofoghi_towards_2016}, influenza \cite{sharpe_evaluating_2016} and mixed diseases \cite{mavragani_infoveillance_2018}.

A considerable number of contributions take into account news events, in particular in relation to severe environmental phenomena such as fire and haze \cite{kibanov_mining_2017, lever_sentimental_2022}, heatwaves \cite{young_social_2021}, flooding \cite{karmegam_spatio-temporal_2020}, rainstorms \cite{wang_spatiotemporal_2019} and earthquakes
\cite{hashimoto_analyzing_2021}. Other commonly explored news events are related to violent actions \cite{ngo_crowd_2016, singh_analyzing_2019, ibar-alonso_opinion_2022}, famous people lives \cite{mcclellan_using_2017, sajjadi_public_2021}, protests \cite{masias_spatial_2021}, online firestorms \cite{strathern_identifying_2022, strathern_against_2020}, industry actions \cite{fernandes_caina_marble_2014} and large-effect decisions such as Brexit \cite{tasoulis_real_2018}.
The impact of politics on behavior change is not very commonly investigated, even though there are some efforts in such a direction: \citet{amelkin_distance_2019} analyze \rev{shifts in} opinions with respect to politicians and their actions, \citet{makazhanov_predicting_2013, zhang_sentiment-change-driven_2017, chaudhry_sentiment_2021} and \citet{tachaiya_sentistance_2021} investigate the role of US election-related events on behavior change and \citet{tang_detecting_2020} perform similar analyses for New Zealand, while \citet{toyoshima_estimating_2022} takes into account the ascension to the throne in Japan as a political phenomenon with possible impact on behavior.
A minor focus is placed on policy introductions, particularly in relation to offshore drilling and airport security  \cite{jiang_topic_2011}, pornographic content censorship \cite{khattak_look_2014}, abortion \cite{graells-garrido_every_2020}, pro-social behavior in videogames \cite{ ryan_lock_2020}, the Goods and Services Tax in India \cite{singh_smart_2020} and mask mandates during the COVID-19 pandemics \cite{mejova_authority_2023}.
Very niche domains of investigation with respect to behavior change-\rev{triggering} events are particular days throughout the year, such as holidays \cite{rodigruez_dominguez_sensing_2017, redondo_hybrid_2020}, daylight saving \cite{linnell_sleep_2021} and sales events \cite{ibrahim_decoding_2019}, online campaigns linked to social movements \cite{foster_metoo_2022, gomes_ribeiro_analyzing_2022, khan_exploring_2022}, exposure to problematic online content \cite{srinivasan_content_2019, sridhar_estimating_2019, wang_twitter_2022}, individual moments of inner change \cite{kushner_bursts_2020, petruzzellis_relation_2023, balsamo_pursuit_2023}, online badge earning \cite{zhang_crowd_2017, yanovsky_one_2021, hoernle_phantom_2022}, sport \cite{hennig_big_2016, wunderlich_big_2022}, online interaction \cite{ahmed_prediction_2022} and project development processes \cite{licorish_understanding_2014}.
We found a small number of works based on a mix of different kinds of events: policies, movies, sports, terrorist attacks, accidents and politics \cite{naskar_emotion_2020}, emerging events from time series analysis \cite{thelwall_sentiment_2011}, COVID-19, FIFA and NBA events \cite{uthirapathy_predicting_2022}.
Almost all of the works take into account external events as behavior change triggers: the only exceptions are \citet{kushner_bursts_2020} and \citet{petruzzellis_relation_2023}, referring to a cognitive change in the subject, and \citet{balsamo_pursuit_2023}, exploring drug recovery.

Moving to the retrieval of events from online content, the examples of indirect detection are relying on manual annotations \cite{sridhar_estimating_2019,gomes_ribeiro_analyzing_2022}, exploiting regular expressions in text \cite{kushner_bursts_2020, balsamo_pursuit_2023} or basing the analyses on time series anomaly detection \cite{thelwall_sentiment_2011, khattak_look_2014, tsytsarau_dynamics_2014, tang_detecting_2020, rosa_event_2020}.
The majority of the research is based on direct event retrieval: knowledge about the events is already embedded in the collected data as key dates of reference (e.g., in \citet{fernandes_caina_marble_2014, ofoghi_towards_2016} and \citet{karmegam_spatio-temporal_2020}), hashtag tracking \cite{amelkin_distance_2019, strathern_against_2020, masias_spatial_2021, strathern_identifying_2022}, online accounts of interest \cite{ahmed_prediction_2022} and information derived from metadata in terms of status recognition \cite{zhang_crowd_2017, yanovsky_one_2021, hoernle_phantom_2022}, exposure to problematic content \cite{srinivasan_content_2019, wang_twitter_2022}, project development state \cite{licorish_understanding_2014} and user-stated opinion change \cite{petruzzellis_relation_2023}.

\subsection{Behavior} \label{subsec:behavior}

\subsubsection{Executive summary}

\begin{figure}[htbp]
  \centering
  \savebox{\largestimage}{\includegraphics[height=.15\textheight]{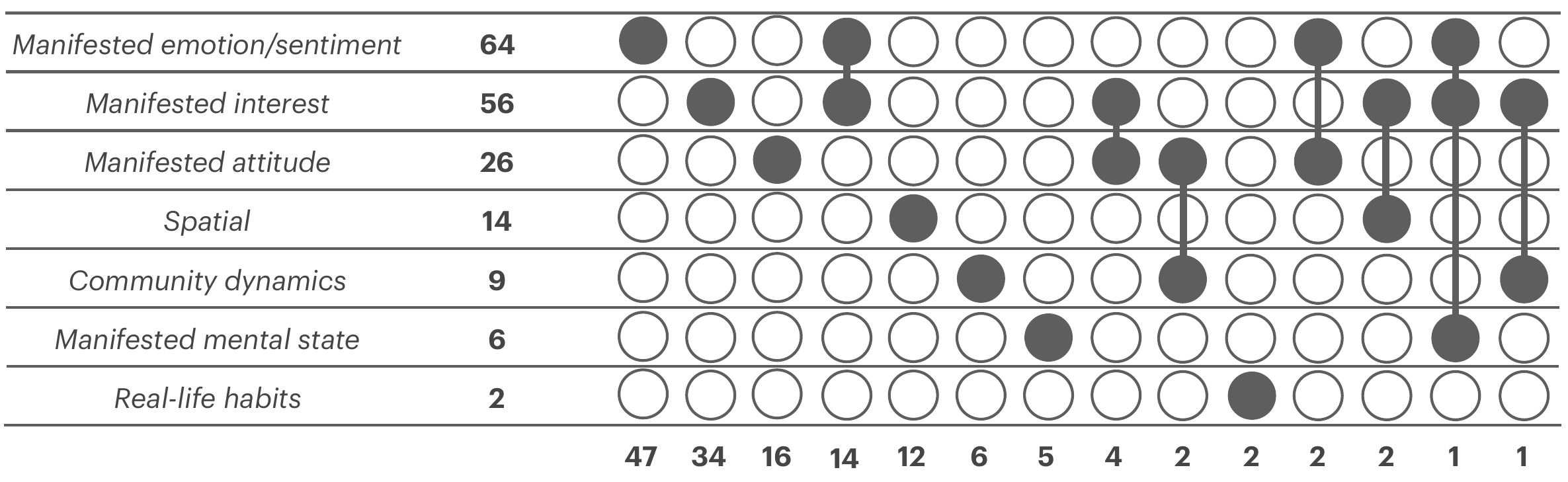}}%
    \begin{subfigure}[b]{0.53\textwidth}
    \centering
    \usebox{\largestimage}
    \caption{Behavior type.}
    \label{fig:behavior_type}
  \end{subfigure}
  \hspace{1.5cm}
  \begin{subfigure}[b]{0.34\textwidth}
    \centering
    \raisebox{\dimexpr.5\ht\largestimage-.5\height}{%
      \includegraphics[height=.105\textheight]{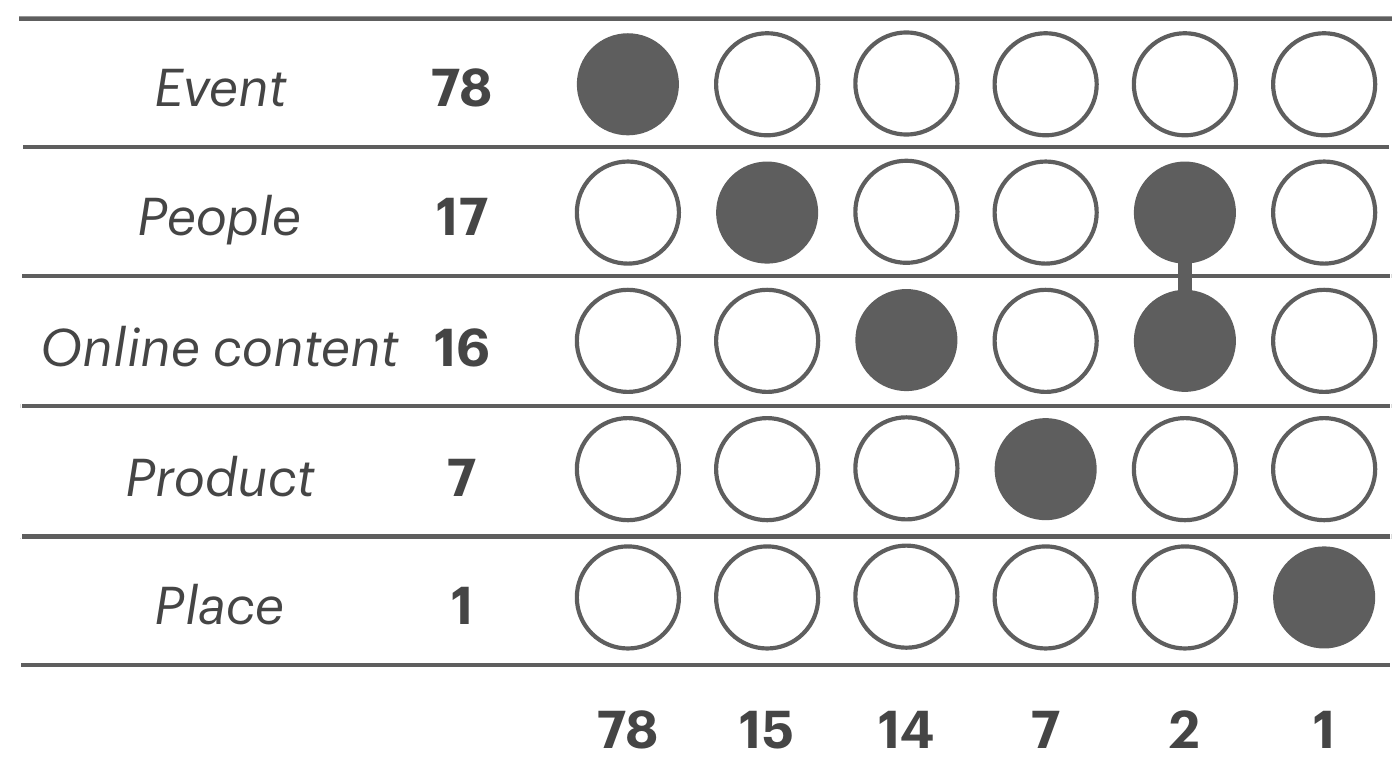}}
    \caption{Object of the behavior.}
    \label{fig:behavior_obj}
  \end{subfigure}\\[4pt]
     \begin{subfigure}[b]{0.98\textwidth}
         \centering
         \includegraphics[width=\textwidth]{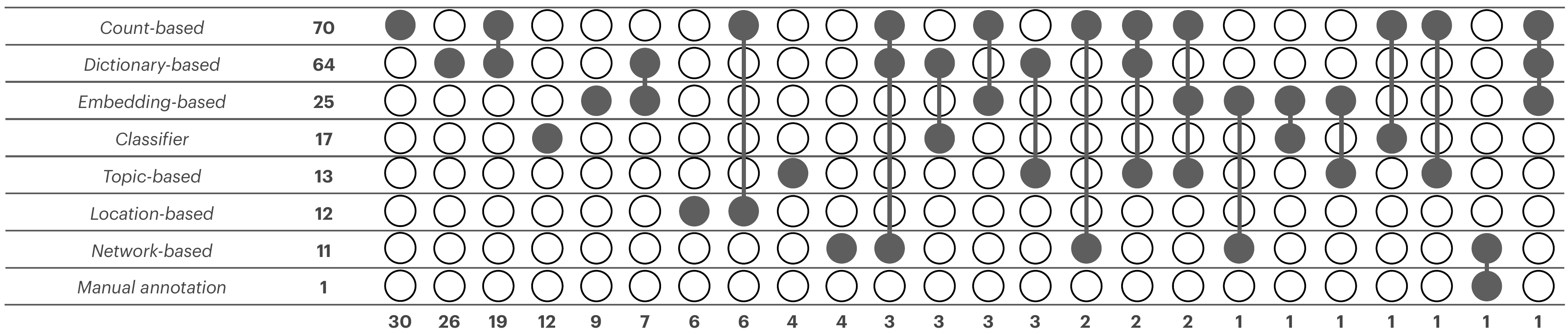}
         \caption{Behavior measure.}
         \label{fig:behavior_proxy}
     \end{subfigure} 
        \caption{\rev{Number of works by behavior type (a), object of the behavior (b), and type of measured used to extract the behavior (c). Dot connectors characterize the simultaneous presence of multiple characteristics.}}
        \label{fig:behavior}
\end{figure}

\begin{figure}[t!]
  \centering
  \includegraphics[width=0.4\linewidth]{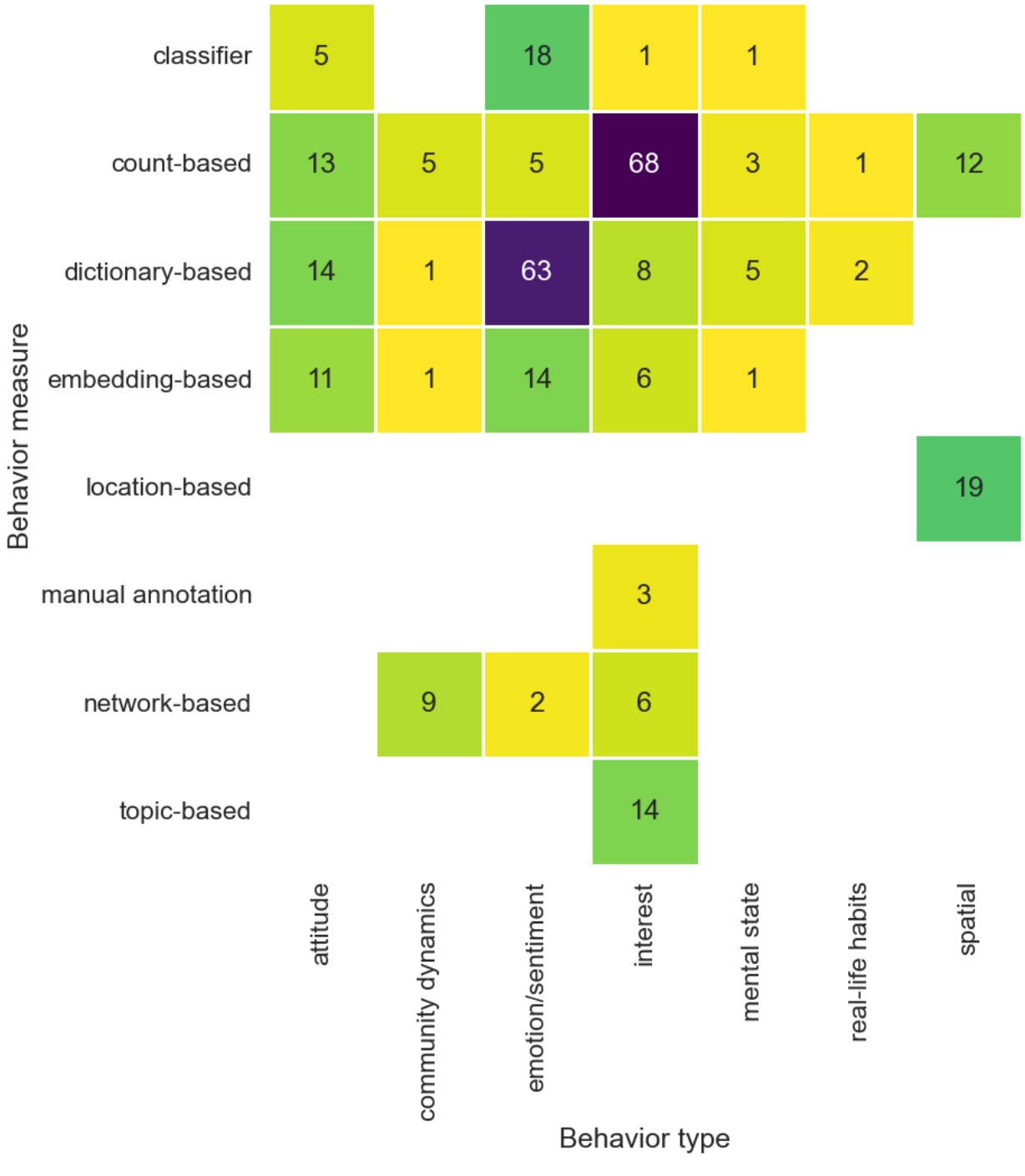}
  \caption{\rev{Number of works by intersection between behavior type and proxy measure used to operationalize the behavior}.}
  \label{fig:behavior_typevsproxy}
  \vspace{-10pt}
\end{figure}

\rev{Given the broad and multidisciplinary lens adopted to classify behaviors in this review, we deal with a wide range of concepts, ranging from over to covert behaviors.}
Approximately 43\% of current research on \rev{behavior} quantification includes the analysis of emotions or sentiment as the primary behavioral reference, at least to some extent (\rev{cf. \Cref{tab:table_behavior_type} and \Cref{fig:behavior_type}}).
Topical interest emerges as the second most frequently studied behavior, featuring in 38\% of the studies.
In 76\% of the research, the behavioral subjects are communities, and behavior is usually tracked by aggregating measurements over many individuals; a few examples of behavior subjects as a collective exist, especially in the case of community dynamics studies.
In the remaining 24\% of the studies, single individuals are the focus of the analysis.
When investigating the objects to which behavior is applied, we find that most behaviors that are explicitly associated with a triggering event \rev{in their change} are applied to an object related to the event itself, usually attitude towards a reference topics (\rev{cf. \Cref{tab:table_behavior_obj} and \Cref{fig:behavior_obj}}).
Notably, nearly 90\% of the research analyzed explores behaviors that are entirely defined online, with no connection with any offline behavior.
Behavior is quantified through various measures (\rev{cf. \Cref{tab:table_behavior_proxy} and \Cref{fig:behavior_proxy}}): 47\% of the studies employ at least one procedure based on simple counting, 43\% use at least one dictionary-based measure, and 17\% use at least one embedding-based tool.
\Cref{fig:behavior_typevsproxy} highlights the narrow range of behavioral measures for some behavior types.
Emotion/sentiment and \rev{expressed} interest, the two most commonly investigated behavior types, are primarily quantified via dictionary-based and count-based measures, respectively.

\subsubsection{Detailed mapping}
The most commonly examined behavior type is linked to the sphere of emotions and sentiment.
Such a concept is often analyzed in terms of generic emotions, linked to a simplified model considering the four basics emotions of anger, fear, happiness and sadness \cite{ngo_crowd_2016} or to more complete lists such as Ekman’s six basic emotions \cite{ofoghi_towards_2016, rosa_event_2020, su_public_2021}, the list defined by Plutchik's emotion wheel \cite{singh_analyzing_2019, karmegam_spatio-temporal_2020, kurten_coronavirus_2021, basile_how_2021, crocamo_surveilling_2021}, the list defined by Russell's model of affect \cite{naskar_emotion_2020} and tailor-made lists expanding basic emotions to more nuanced concepts \cite{cao_learning_2021, pellert_dashboard_2020, iglesias-sanchez_contagion_2020, choudrie_applying_2021, caliskan_how_2022, han_weibo_2022}.
Sentiment is even more commonly explored in terms of positive and negative feelings about given entities, for instance in \citet{tan_interpreting_2014, giachanou_explaining_2016, bui_temporal_2016, hu_systematic_2020} and \citet{alattar_using_2021}, but there are also analyses assessing emotional concern \cite{ortega_modelling_2021, li_public_2022}, moral valence \cite{mejova_authority_2023}, perception \cite{sun_new_2021, jung_study_2021} and polarization \cite{ramaciotti_morales_auditing_2021}.
Emotions and sentiment are often studied in combination with \rev{expressed} interest, particularly in terms of the popularity of topics (such as in \citet{jiang_topic_2011, tasoulis_real_2018, valdez_social_2020, kurten_coronavirus_2021} and \citet{rahmanti_social_2021}), but are also evaluated in connection with interest and mental state \cite{marshall_using_2022}, or attitude \cite{li_dynamic_2022, mejova_authority_2023}.

The second overall most examined behavior is expressed \rev{interest}: the focus is often placed on topical engagement (as, for instance, in \citet{achananuparp_who_2012, sidana_health_2018, shi_social_2020, alzamzami_monitoring_2021} and \citet{abramova_collective_2022}) but there are a prominent number of works exploring online learning behavior, particularly from 2020 as an effect of the COVID-19 pandemic \cite{peng_topic_2020, zhang_measuring_2021, dascalu_before_2021, zhao_academic_2021, dermy_dynamic_2022, tao_data_2022}. 
Other forms of interest expressed online are considered in terms of sharing and posting behaviors \cite{licorish_understanding_2014, basile_how_2021, mejova_youtubing_2021}, contribution to Q\&A websites \cite{yanovsky_one_2021, hoernle_phantom_2022}, search behavior \cite{khattak_look_2014, epure_modeling_2017, mavragani_infoveillance_2018, sajjadi_public_2021} and generic browsing activity, such as in \citet{denecke_how_2013, sharpe_evaluating_2016, link_estimating_2017, jindal_intend_2018} and \citet{wang_twitter_2022}.
As already described, interest is often analyzed in relation to emotions, sentiment and mental state, but there are also some examples of studies linked to attitude \cite{licorish_understanding_2014, low_natural_2020, gomes_ribeiro_analyzing_2022, balsamo_pursuit_2023}, spatial behaviors \cite{kibanov_mining_2017, mejova_youtubing_2021} and community dynamics \cite{scherz_whatsapp_2022}.

The attitude of online users is analyzed to a minor extent, as a generic concept referring to how people show themselves online \cite{joshi_analysis_2020, kushner_bursts_2020, low_natural_2020, harywanto_bertweet-based_2022} or, more specifically, in terms of tone of voice (such as in \citet{licorish_understanding_2014, flekova_exploring_2016, sciandra_covid-19_2020, ryan_lock_2020} and \citet{balsamo_pursuit_2023}), stance or preference \cite{makazhanov_predicting_2013, graells-garrido_every_2020, tachaiya_sentistance_2021, dinh_computational_2022, mejova_authority_2023}, mental associations \cite{tahmasbi_go_2021}, policy acceptance \cite{li_dynamic_2022}, social identity \cite{rogers_using_2021} and toxicity \cite{almerekhi_investigating_2022}.
Attitude is examined on its own or in relation to interest, emotion/sentiment, or community dynamics \cite{strathern_against_2020, strathern_identifying_2022}.
Spatial behaviors constitute a small subset of the overall focus and are referred to generic human mobility patterns \cite{kibanov_mining_2017, lam_detecting_2019, shepherd_domestic_2021, mejova_youtubing_2021, aljeri_big_2022}, crowd movement \cite{nayak_exploring_2015, franca_visualizing_2016, rodigruez_dominguez_sensing_2017, santa_statistical_2019, redondo_hybrid_2020, toyoshima_estimating_2022}, tourism-related mobility \cite{liu_categorization_2020, jing_fine-grained_2020} and spatial density \cite{masias_spatial_2021}.
Online community dynamics is also examined to a limited extent, specifically in relation to the interaction of users in a communication network \cite{uddin_impact_2014}, users' cooperation on collaborative crowdsourcing sites \cite{zhang_crowd_2017}, group and role dynamics \cite{zygmunt_achieving_2020, strathern_against_2020}, community evolution \cite{hu_detecting_2020, tang_detecting_2020, karatas_novel_2022, strathern_identifying_2022} and engagement and network structure \cite{scherz_whatsapp_2022}.
Finally, a small number of works analyze mental state \cite{li_using_2015, vioules_detection_2018, li_modeling_2020, stevens_desensitization_2021, ren_psychological_2022, marshall_using_2022}, with a focus on stress, depression and anxiety, and real-life habits of subjects active online, with a particular emphasis on alcohol drinking \cite{liu_assessing_2017} and sleep \cite{linnell_sleep_2021} behaviors.

The subject of reference for the behaviors to be analyzed is usually a group of users, as data is often treated in an aggregated fashion (such as in \citet{mcclellan_using_2017, kaur_monitoring_2020, evers_covid-19_2021, tahmasbi_go_2021, toyoshima_estimating_2022}), but there are several examples of focus placed on individual behaviors. 
For instance, \citet{makazhanov_predicting_2013, zhao_explainable_2016, naskar_emotion_2020, zhao_academic_2021} and \citet{ahmed_prediction_2022} analyze behaviors at a single-user level and \citet{licorish_understanding_2014, kim_analysis_2021, aljeri_big_2022, hoernle_phantom_2022} and \citet{scherz_whatsapp_2022} are examples of a minor research line exploring a combination of individuals and groups as behavior subjects.

Provided that we always take into account a human subject as the reference for our analyzed behaviors, we are interested in understanding what are the most common objects to which the behaviors of reference are applied.
Not surprisingly, most of the behaviors that are explicitly associated with a triggering event \rev{in relation to their change} are applied to an object related to the event itself, such as emerging events of different types detected as time series anomalies \cite{thelwall_sentiment_2011}, disease-related events \cite{ofoghi_towards_2016, yeung_face_2020, evers_covid-19_2021} and violent phenomena \cite{singh_analyzing_2019}.
A considerable number of works define people as objects of the behavior: such works interpret the behavior subject also as the object of reference (i.e., self-applied behavior) \cite{rogers_using_2021, ren_psychological_2022} or refer to external individuals as objects (such as done by \citet{makazhanov_predicting_2013, giachanou_explaining_2016, zhang_sentiment-change-driven_2017, joshi_analysis_2020} and  \citet{ramaciotti_morales_auditing_2021}).
Other research lines take into account online content (as in \citet{khattak_look_2014, sridhar_estimating_2019,  zhang_measuring_2021, tao_data_2022} and \citet{petruzzellis_relation_2023}), commercial products \cite{wang_anomaly_2015, ibrahim_decoding_2019, kim_analysis_2021, li_dynamic_2022, ahmed_prediction_2022, steinke_sentiment_2022, rabiu_drift_2023} and places \cite{liu_categorization_2020} as behavior objects. 

Almost all of the existing research explores behaviors originally defined online: there are only a small number of papers focusing on online data to study offline behaviors, in particular referring to spatial mobility (such as in \citet{santa_statistical_2019, lam_detecting_2019} and \citet{shepherd_domestic_2021}) or to real-life habits \cite{liu_assessing_2017, linnell_sleep_2021}.
\citet{kibanov_mining_2017} is an interesting case of a mixed source paper describing both interest-related and spatial behaviors.

The most common type of measure for quantifying behaviors are count-based procedures, mostly assessed in relation to interest as the behavior type of reference (such as in \citet{epure_modeling_2017, kurten_coronavirus_2021, rahmanti_social_2021, young_social_2021} and \citet{erokhin_covid-19_2022}) but also in the context of community dynamics \cite{zhang_crowd_2017, zygmunt_achieving_2020, strathern_against_2020, strathern_identifying_2022, karatas_novel_2022}, emotion/sentiment \cite{amelkin_distance_2019, sun_new_2021, ortega_modelling_2021, jung_study_2021}, spatial behaviors \cite{santa_statistical_2019, liu_categorization_2020, jing_fine-grained_2020, shepherd_domestic_2021, masias_spatial_2021, mejova_youtubing_2021, toyoshima_estimating_2022}, attitude (as in \citet{makazhanov_predicting_2013, flekova_exploring_2016, joshi_analysis_2020,  sciandra_covid-19_2020, dinh_computational_2022} and  \citet{mejova_authority_2023}), real-life habits \cite{linnell_sleep_2021} and mental state \cite{vioules_detection_2018, marshall_using_2022}.
The most commonly employed count-based measures include the evaluation of the number of posts published by users (such as in \citet{achananuparp_who_2012, mcclellan_using_2017, wang_spatiotemporal_2019, rahmanti_social_2021}), the number of textual elements (words, hashtags, and emojis) shared (such as in \citet{flekova_exploring_2016,  srinivasan_content_2019, kushner_bursts_2020, low_natural_2020} and \citet{wunderlich_big_2022}), the number of community members performing certain actions \cite{licorish_understanding_2014, shepherd_domestic_2021, karatas_novel_2022, petruzzellis_relation_2023}, engagement metrics such as the number of followers and likes to posts \cite{liu_personalized_2010, vioules_detection_2018, kushner_bursts_2020, gomes_ribeiro_analyzing_2022} and the number of check-ins websites \cite{liu_categorization_2020, shi_temporal_2020}.
Another common behavior measure type are dictionary-based procedures, referring to pre-defined dictionary sets or keyword lists and mainly related to emotion and sentiment as behaviors of reference, such as in \citet{fernandes_caina_marble_2014, hu_systematic_2020, yeung_face_2020, ibar-alonso_opinion_2022} and \citet{lwin_evolution_2022}.
Dictionary-based measures are also employed to analyze attitude, such as in \citet{ryan_lock_2020, cao_learning_2021, rogers_using_2021, foster_metoo_2022} and
\citet{gomes_ribeiro_analyzing_2022}, interest \cite{zhao_academic_2021, tao_data_2022, wang_twitter_2022, balsamo_pursuit_2023}, mental state \cite{li_using_2015, vioules_detection_2018, stevens_desensitization_2021, ren_psychological_2022}, community-dynamics \cite{zhang_crowd_2017} and real-life habits \cite{liu_assessing_2017}.
Embedding-based behavior measures are used to quantify emotions and sentiment, such as in \citet{ofoghi_towards_2016, yu_temporal_2020, kahanek_temporal_2021, caliskan_how_2022, lwin_evolution_2022} and \citet{han_weibo_2022}, attitude, for instance in \citet{sridhar_estimating_2019, joshi_analysis_2020, tahmasbi_go_2021, tachaiya_sentistance_2021, harywanto_bertweet-based_2022} and \citet{almerekhi_investigating_2022}, interest \cite{almerekhi_are_2020, liu_monitoring_2021, lu_agenda-setting_2022, petruzzellis_relation_2023}, community dynamics \cite{tang_detecting_2020} and mental state \cite{li_modeling_2020}.
Such measures include both word frequency-based embeddings, such as in \citet{ofoghi_towards_2016} and \citet{sridhar_estimating_2019}, and word embedding models, such as in \citet{tahmasbi_go_2021} and \citet{balsamo_pursuit_2023}.
Classifier-based behavior quantification measures are also explored, mainly in the context of emotions and sentiment (such as in \citet{tan_interpreting_2014, hennig_big_2016, bui_temporal_2016, steinke_sentiment_2022}) but also in relation to attitude \cite{graells-garrido_every_2020, li_dynamic_2022, almerekhi_investigating_2022}, interest \cite{alzamzami_monitoring_2021} and mental state \cite{li_using_2015}.
Topic-based measures are related to topic modeling techniques evaluated in the context of interest as the behavior of reference, such as in \citet{zhao_explainable_2016, sidana_health_2018, peng_topic_2020, liu_monitoring_2021}, location-based measures are linked to spatial behaviors, for instance in \citet{nayak_exploring_2015, franca_visualizing_2016, lam_detecting_2019, toyoshima_estimating_2022} and network-based measures are used to analyze community dynamics \cite{uddin_impact_2014, zygmunt_achieving_2020, strathern_against_2020, hu_detecting_2020, tang_detecting_2020, strathern_identifying_2022, scherz_whatsapp_2022}, emotions and sentiment \cite{jung_study_2021, ramaciotti_morales_auditing_2021} and interest \cite{hashimoto_analyzing_2021, dascalu_before_2021}.
Finally, manual annotation to quantify behavior can be included in addition to other types of measures \cite{scherz_whatsapp_2022}.

A vast number of works focus on a single behavior measure type, but a consistent share takes into account combinations of measures: the most common patterns are count-based and dictionary-based (such as in \citet{zhang_crowd_2017, amelkin_distance_2019, zhao_academic_2021} and \citet{ortega_modelling_2021}), dictionary-based and embedding-based \cite{ofoghi_towards_2016, ngo_crowd_2016, sridhar_estimating_2019, almerekhi_are_2020, cao_learning_2021, crocamo_surveilling_2021, lwin_evolution_2022}, count-based and location-based \cite{kibanov_mining_2017, jing_fine-grained_2020, shepherd_domestic_2021, masias_spatial_2021, mejova_youtubing_2021, toyoshima_estimating_2022}, count-based, dictionary-based and network-based \cite{strathern_against_2020, jung_study_2021, strathern_identifying_2022}, dictionary-based and classifier-based \cite{tan_interpreting_2014, wang_anomaly_2015, li_using_2015}, count-based and embedding-based \cite{joshi_analysis_2020, lu_agenda-setting_2022, petruzzellis_relation_2023} or dictionary-based and topic-based \cite{tasoulis_real_2018, flocco_analysis_2021, zhou_examination_2021}.
The following combinations of behavior quantification proxies are less commonly explored: count-based and network-based \cite{zygmunt_achieving_2020, dascalu_before_2021}, count-based, dictionary-based and topic-based \cite{low_natural_2020, valdez_social_2020}, count-based, embedding-based and topic-based \cite{marshall_using_2022, han_weibo_2022}, embedding-based and network-based \cite{tang_detecting_2020}, embedding-based and classifier \cite{almerekhi_investigating_2022}, embedding-based and topic-based \cite{liu_monitoring_2021}, count-based and classifier \cite{rahmanti_social_2021}, count-based and topic-based \cite{shi_social_2020}, count-based, dictionary-based and embedding-based \cite{balsamo_pursuit_2023}.

\subsection{\rev{Behavior Change}} \label{subsec:behavior_change}
\subsubsection{Executive summary}

\begin{figure}[!htbp]
  \centering
       \begin{subfigure}[b]{0.98\textwidth}
         \centering
         \includegraphics[width=\textwidth]{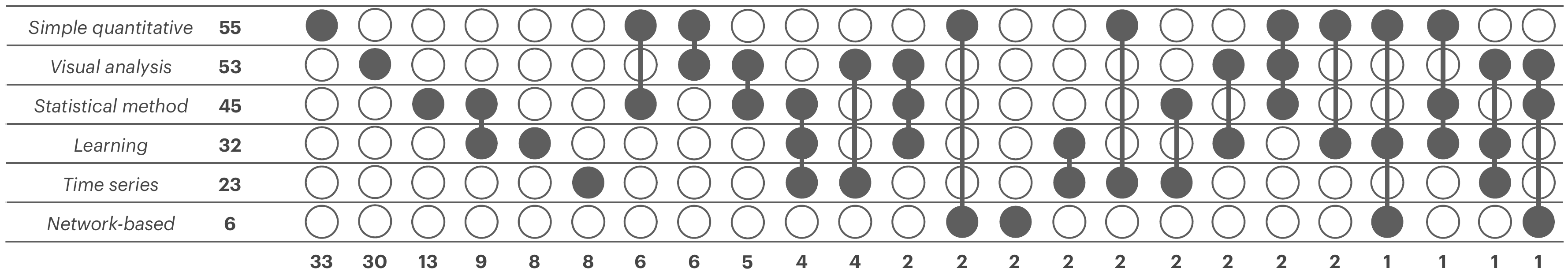}
         \caption{Behavior change techniques.}
         \label{fig:change_tech}
     \end{subfigure} \\[10pt]
    \begin{subfigure}[b]{0.80\textwidth}
         \centering
         \includegraphics[width=\textwidth]{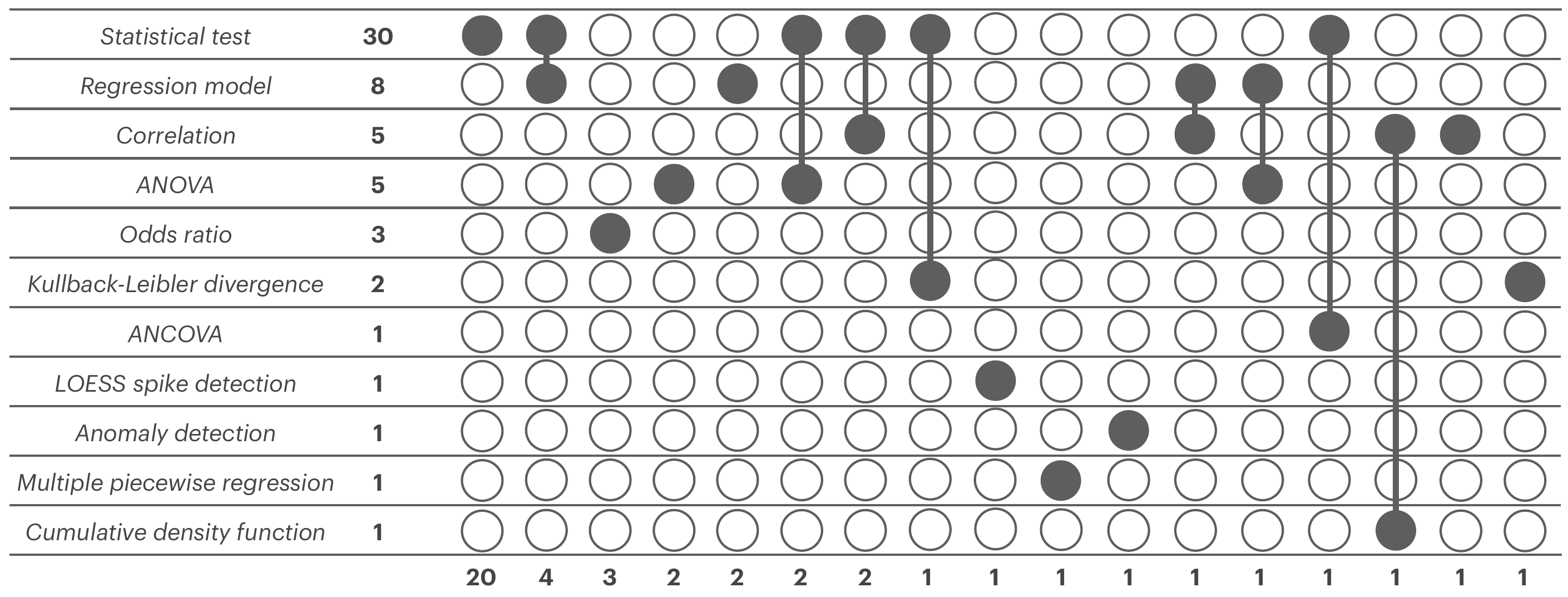}
         \caption{Statistical techniques.}
         \label{fig:change_stat}
     \end{subfigure}\\[10pt]
    \begin{subfigure}[b]{0.63\textwidth}
         \centering
         \includegraphics[width=\textwidth]{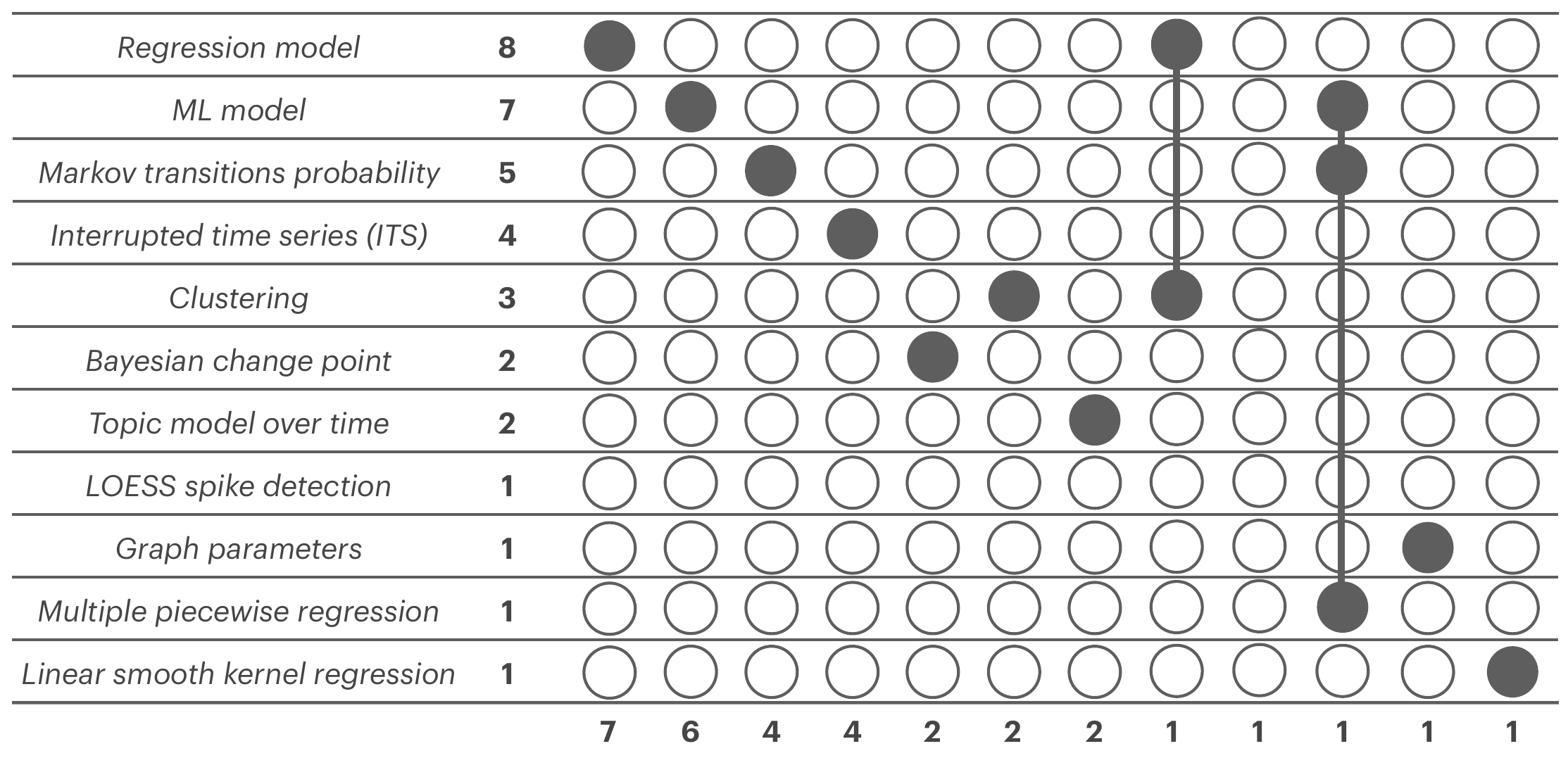}
         \caption{Model learning techniques.}
         \label{fig:change_learn}
     \end{subfigure}\\[10pt]
  \savebox{\largestimage}{\includegraphics[height=.14\textheight]{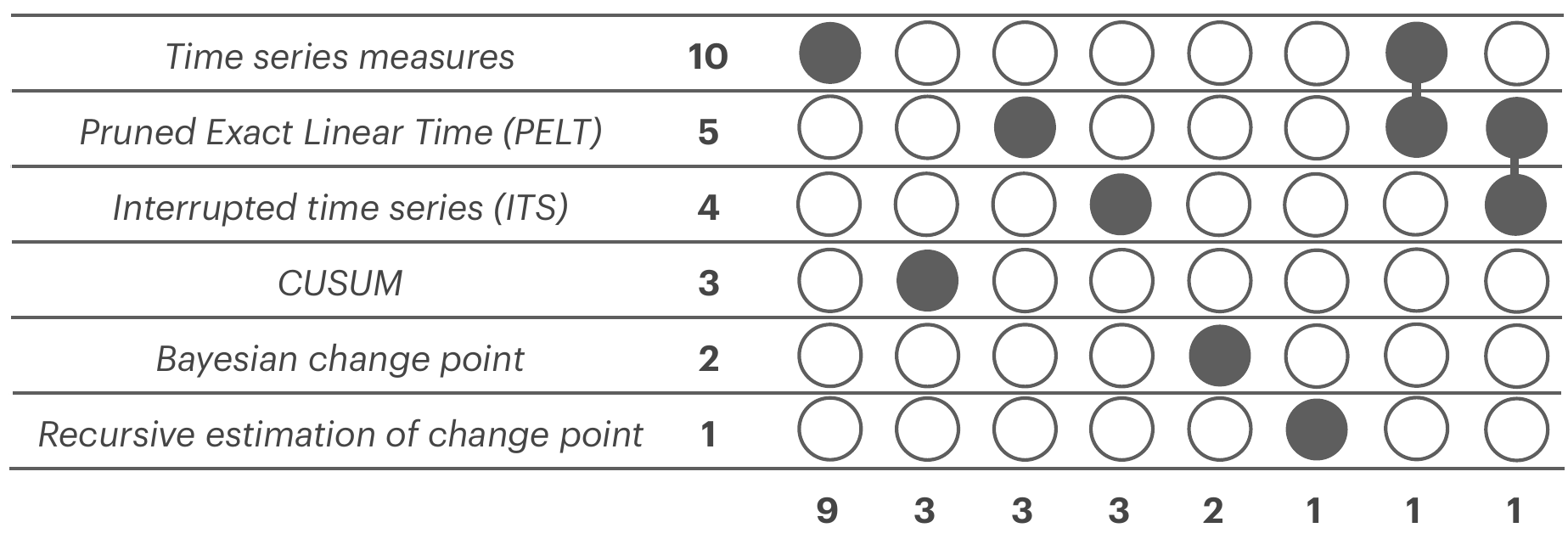}}%
  \begin{subfigure}[b]{0.4\textwidth}
    \centering
    \raisebox{\dimexpr.5\ht\largestimage-.5\height}{%
      \includegraphics[height=.11\textheight]{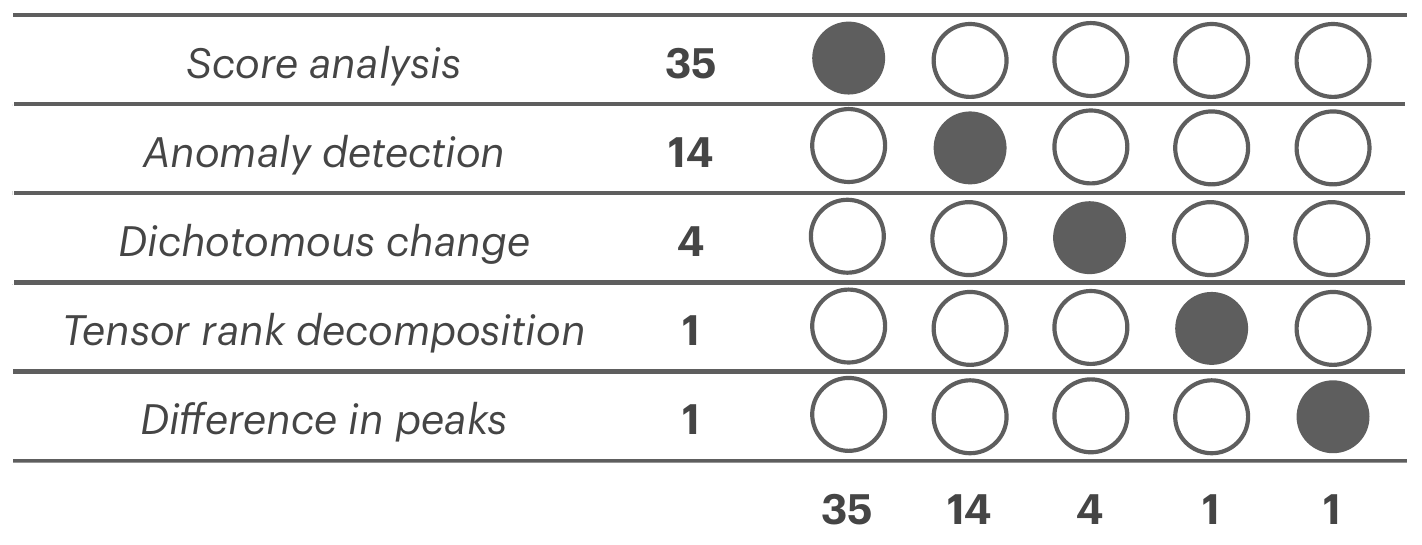}}
    \caption{Quantitative techniques.}
    \label{fig:change_quant}
  \end{subfigure}
  \begin{subfigure}[b]{0.55\textwidth}
    \centering
    \usebox{\largestimage}
    \caption{Time series techniques.}
    \label{fig:timeseries}
  \end{subfigure}
        \caption{\rev{Number of works by behavior change detection technique (a) and details on the specific statistical (b), model learning-based (c), quantitative (d), and time series (e) methods used. Dot connectors characterize the simultaneous presence of multiple characteristics.}}
        \label{fig:behavior_change}
\end{figure}

\rev{The second main aspect of interest of the current review is the mapping of measures used to analyze behavior shifts.}
Around 73\% of the methods rely either on the direct comparison between \rev{measured values of a behavior} taken at different points in time (at least partially in 37.2\% of the total works) or on a simple visual assessment of temporal trends of the target \rev{behavior measure} (at least partially in 35.8\% of the research, see \rev{\Cref{tab:table_behavior_change_tech} and \Cref{fig:change_tech}}).
As depicted in \Cref{fig:change_quant}, the most common quantitative technique for analyzing behavior change is score comparison (63.6\% of the studies considering quantitative methods), which simply measures behavior differences over time with or without explicitly calculating a measure of change.
Among the remaining types of studies, statistical methods are quite popular (they are considered, at least to some extent, in 30.4\% of the research): 66.7\% of such works use statistical tests to measure significant shifts in the target behavior (\Cref{fig:change_stat}).
A consistent 21.6\% of the total works venture, even partially, into model learning using a variety of tools including linear regression, machine learning models, Markov chains, and interrupted time series (\Cref{fig:change_learn}).
Techniques based on time series mainly touch upon structural properties, time series modeling, or anomaly detection. Network-based procedures receive generally less attention.

\rev{Given that behavior is interpreted as an overt or covert manifestation by individuals or groups and behavior change as a shift in such behaviors, the} subjects of behavior change can be different from the subjects of behaviors.
\rev{In fact,} the focus can be placed on different kinds of aggregation when shifting from the description of a behavior to the study of its change \rev{(e.g., behavior adopted by an individual, change considered at a group level after the aggregation of multiple individual behaviors)}.
The primary subject \rev{of behavior change is} defined at a group level in nearly 80\% of the studies.
Behavior change is described as a discrete phenomenon in 45\% of the studies, and as a gradual process in 40\% of them; the remaining 15\% consider both discrete and gradual changes.
When looking at the temporal characteristics of behavior change, we found that nearly 68\% of the gradual behavior changes are defined in a continuous fashion: more than two time bins are compared, and the evolution over time can be considered continuous or can be approximated as such.
The remaining research on gradual change partially considers a gradual behavior change over a small number of phases.

\subsubsection{Detailed mapping}
In terms of quantitative \rev{behavior} change detection techniques, a considerable number of works rely on the analysis of numerical scores, by computing a measure of behavior change between time bins, such as in \citet{liu_personalized_2010, jiang_modeling_2015, sidana_health_2018, sridhar_estimating_2019, hu_systematic_2020} and \citet{tachaiya_sentistance_2021}, or by comparing behavior scores over time, such as in \citet{uddin_impact_2014, li_using_2015, pellert_dashboard_2020, cao_learning_2021} and \citet{shepherd_domestic_2021}.
Anomaly detection is another popular quantitative change detection technique: by comparing the observed behavior to a ground truth \cite{nayak_exploring_2015, rodigruez_dominguez_sensing_2017, amelkin_distance_2019, redondo_hybrid_2020} or to numerical thresholds, defined both manually \cite{jiang_topic_2011, achananuparp_who_2012, tan_interpreting_2014, kibanov_mining_2017, lam_detecting_2019, alattar_using_2021, rabiu_drift_2023} or via automatic methods \cite{jiang_topic_2011, joshi_analysis_2020, almerekhi_are_2020}, it is possible to detect unusual characteristics \rev{and shifts} of the behavior \rev{itself}.
More trivial techniques, such as the assessment of change as a boolean parameter, are also taken into account \cite{makazhanov_predicting_2013, zygmunt_achieving_2020, yanovsky_one_2021, almerekhi_investigating_2022}: the behavior change is, in this case, detected as the existence of a difference between two behavioral states.
Finally, some less commonly explored quantitative detection techniques are tensor rank decomposition \cite{flocco_analysis_2021} and difference in behavioral curves peaks \cite{linnell_sleep_2021}.
Visual \rev{comparison and analysis} of \rev{subsequent} behaviors is the second most commonly analyzed change detection technique: behavior change is assessed in terms of comparison of heat maps defined over spatial activity data \cite{franca_visualizing_2016, mavragani_infoveillance_2018, li_modeling_2020, sun_new_2021, aljeri_big_2022, lever_sentimental_2022} or semantic features of language \cite{dascalu_before_2021}, visual dynamics of a trend over time (such as in \citet{hennig_big_2016, flekova_exploring_2016, hu_systematic_2020, basile_how_2021} and \citet{erokhin_covid-19_2022}), analysis of topic evolution (such as in \citet{tasoulis_real_2018, valdez_social_2020, alzamzami_monitoring_2021, liu_monitoring_2021, mejova_youtubing_2021} and \citet{marshall_using_2022}), emotion plots change \cite{choudrie_applying_2021, basile_how_2021} and pattern frequencies \cite{aljeri_big_2022}.
Another change detection technique family that covers a consistent share of the overall research is that of statistical methods: a vast majority of the works considering techniques part of this group tend to focus on statistical tests, such as z-score test \cite{ren_psychological_2022}, chi-square test \cite{licorish_understanding_2014, kim_analysis_2021, scherz_whatsapp_2022}, Kruskall-Wallis H test \cite{iglesias-sanchez_contagion_2020}, Tukey's range test \cite{karmegam_spatio-temporal_2020} and t-test \cite{ofoghi_towards_2016, wang_twitter_2022}. 
However, there are some examples of change detection based on regression models, such as \citet{mavragani_infoveillance_2018, santa_statistical_2019, wang_spatiotemporal_2019, masias_spatial_2021} and \citet{wunderlich_big_2022}, correlation analysis \cite{kibanov_mining_2017, mavragani_infoveillance_2018, sajjadi_public_2021, young_social_2021, li_public_2022}, ANOVA \cite{licorish_understanding_2014, liu_categorization_2020, ryan_lock_2020, su_public_2021, foster_metoo_2022}, odds ratios \cite{sciandra_covid-19_2020, stevens_desensitization_2021, petruzzellis_relation_2023}, Kullback-Leibler divergence \cite{ofoghi_towards_2016, peng_topic_2020}, ANCOVA \cite{zhang_measuring_2021}, LOESS spike regression \cite{giachanou_explaining_2016}, multiple piecewise regression \cite{zhao_academic_2021}, cumulative density function \cite{kibanov_mining_2017} and anomaly detection \cite{lam_detecting_2019}.
Note that this latter sub-type of statistical method-based change detection is a particular type of "anomaly detection" defined as a quantitative technique, in which a statistical significance assessment is performed.
Change detection techniques based on parameters or models learning (cf. \Cref{fig:change_learn}) are mainly based on simple regression models (for instance, models reported in \citet{mavragani_infoveillance_2018, ryan_lock_2020, karmegam_spatio-temporal_2020, low_natural_2020} and \citet{wunderlich_big_2022}), broadly defined machine learning techniques (such as in \citet{liu_assessing_2017, harywanto_bertweet-based_2022, ahmed_prediction_2022, hoernle_phantom_2022} and \citet{petruzzellis_relation_2023}), Markov models \cite{bui_temporal_2016, link_estimating_2017, epure_modeling_2017, naskar_emotion_2020, zhao_academic_2021} and Interrupted Time Series (ITS) \cite{srinivasan_content_2019, mejova_youtubing_2021, mejova_authority_2023, balsamo_pursuit_2023}.
Other learning methods that are examined, even though to a minor extent, are clustering \cite{santa_statistical_2019, shi_temporal_2020, uthirapathy_predicting_2022}, Bayesian change point analysis \cite{sharpe_evaluating_2016, singh_analyzing_2019}, topic modeling \cite{zhao_explainable_2016, peng_topic_2020}, LOESS spike detection \cite{giachanou_explaining_2016}, graph parameters \cite{uddin_impact_2014}, multiple piecewise regression \cite{zhao_academic_2021} and linear smooth kernel regression \cite{iglesias-sanchez_contagion_2020}.

Time series-based change detection techniques are generally focused on structural properties and anomalies in time series (cf. \Cref{fig:timeseries}), for instance in \citet{nayak_exploring_2015, mcclellan_using_2017, ibrahim_decoding_2019, sajjadi_public_2021} and \citet{leon-sandoval_monitoring_2022}, but there are also some works examining the Pruned Exact Linear Time (PELT) algorithm \cite{strathern_against_2020, yeung_face_2020, valdez_social_2020, flocco_analysis_2021, mejova_youtubing_2021}, ITS \cite{srinivasan_content_2019, mejova_youtubing_2021, mejova_authority_2023, balsamo_pursuit_2023}, CUmulative SUM control chart (CUSUM) \cite{denecke_how_2013, tasoulis_real_2018, khan_exploring_2022}, Bayesian change point analysis \cite{sharpe_evaluating_2016, singh_analyzing_2019} and recursive estimation of change point \cite{theodosiadou_change_2021}.

Finally, network-based behavior change detection techniques are examined to a very small extent and particularly in the form of network comparison \cite{zygmunt_achieving_2020, hu_detecting_2020, dascalu_before_2021, karatas_novel_2022}, graph parameters~\cite{uddin_impact_2014} and graph anomaly detection~\cite{amelkin_distance_2019}.

While the majority of research tends to focus on a single change detection technique family, there are some examples of combinations of multiple ones, especially because some change assessment techniques can be affiliated with several families at the same time.
This is the case, for instance, of works taking into account regression models~\cite{santa_statistical_2019, wang_spatiotemporal_2019, ryan_lock_2020, karmegam_spatio-temporal_2020, masias_spatial_2021} and LOESS spike detection~\cite{giachanou_explaining_2016} to quantify behavior change: such change detection techniques can be classified both as statistical and learning-based methods.
Conversely, some works combine statistical and learning-based families by taking completely different methods into account: for instance, clustering and regression models \cite{santa_statistical_2019}, topic modeling and Kullback-Leibler divergence~\cite{peng_topic_2020}, and machine learning models and odds ratios~\cite{petruzzellis_relation_2023}.
Other combinations of change detection techniques families that are often explored are visual analysis and statistical methods~\cite{young_social_2021, zhang_measuring_2021, foster_metoo_2022, strathern_identifying_2022, li_public_2022}, quantitative methods and statistical ones~\cite{licorish_understanding_2014, ofoghi_towards_2016, kibanov_mining_2017, lam_detecting_2019, liu_categorization_2020, steinke_sentiment_2022}, quantitative methods and visual analysis~\cite{hu_systematic_2020, lu_agenda-setting_2022, aljeri_big_2022, almerekhi_investigating_2022, ibar-alonso_opinion_2022, lever_sentimental_2022}, statistical methods, parameters learning and time series~\cite{srinivasan_content_2019, zhao_academic_2021, mejova_authority_2023, balsamo_pursuit_2023} and visual analysis and time series~\cite{tasoulis_real_2018, ibrahim_decoding_2019, strathern_against_2020, valdez_social_2020}.
A small share of research explores the intersection between visual analysis, statistical methods and learning~\cite{mavragani_infoveillance_2018, wunderlich_big_2022}, quantitative-based and time series methods~\cite{nayak_exploring_2015, flocco_analysis_2021}, statistical methods and time series~\cite{sajjadi_public_2021, gomes_ribeiro_analyzing_2022}, visual analysis and learning~\cite{liu_assessing_2017, hoernle_phantom_2022}, quantitative-based methods, visual analysis and statistical methods~\cite{almerekhi_investigating_2022, scherz_whatsapp_2022}, quantitative-based and learning~\cite{yanovsky_one_2021, ahmed_prediction_2022}, quantitative-based, learning and network-based techniques~\cite{uddin_impact_2014}, quantitative methods, statistical methods and learning~\cite{low_natural_2020}, visual analysis, learning and time series~\cite{mejova_youtubing_2021} and visual analysis, statistical methods and network-based techniques~\cite{dascalu_before_2021}.

\begin{figure}[ht!]
  \centering
  \savebox{\largestimage}{\includegraphics[height=.32\textheight]{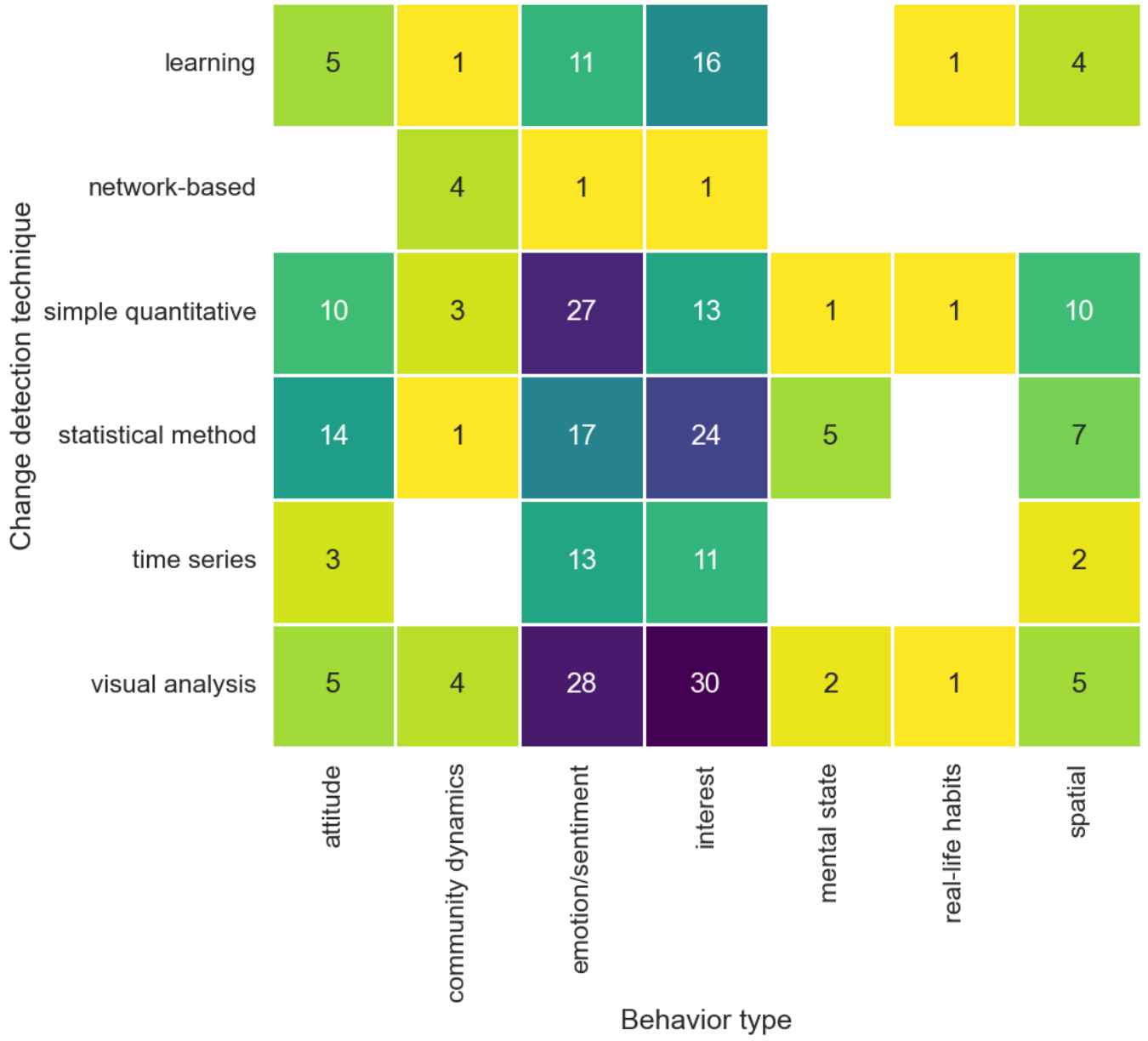}}%
  \begin{subfigure}[b]{0.45\textwidth}
    \centering
    \usebox{\largestimage}
    \caption{Behavior type and change technique.}
    \label{fig:behavior_tech_heat}
  \end{subfigure}
  \hspace{0.5cm}
  \begin{subfigure}[b]{0.47\textwidth}
    \centering
    \raisebox{\dimexpr.5\ht\largestimage-.5\height}{%
      \includegraphics[height=.29\textheight]{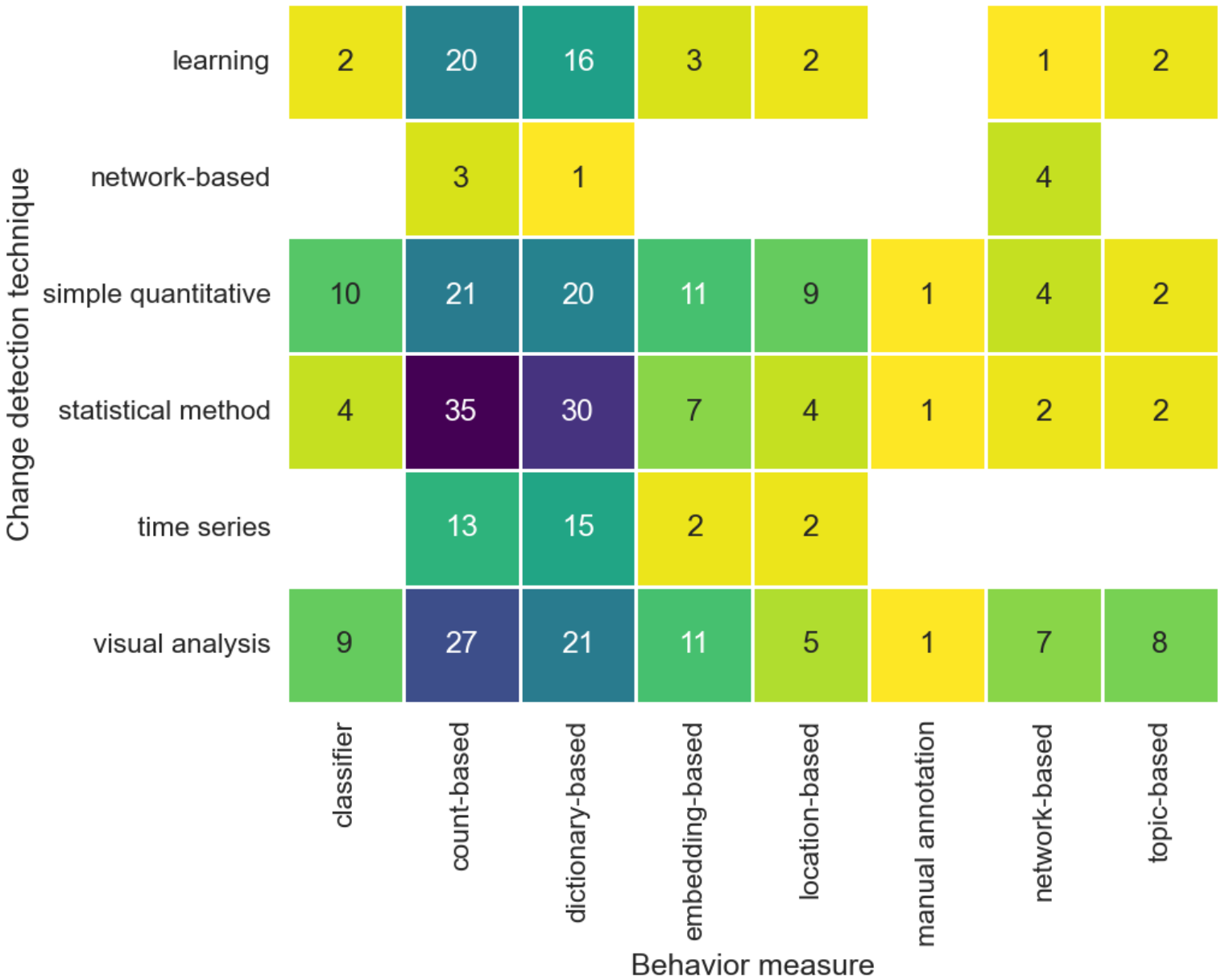}}
    \caption{Behavior measure type and change technique.}
    \label{fig:proxy_tech_heat}
  \end{subfigure}
  \caption{\rev{Number of works by intersection between behavior change detection technique and behavior type (a), and between behavior change detection technique and behavior measure (b).}}
\end{figure}

We analyze existing research patterns in terms of the evaluation of behavior types and related change detection techniques (\Cref{fig:behavior_tech_heat}).
Network-based and time series-based techniques are overall less explored, but time series-based methods are often employed on emotion/sentiment, such as in \citet{tsytsarau_dynamics_2014, yeung_face_2020} and \citet{leon-sandoval_monitoring_2022}, and interest, such as in \citet{sharpe_evaluating_2016, theodosiadou_change_2021,sajjadi_public_2021}.
Moreover, parameter learning is consistently taken into account to measure interest change, such as in \citet{sharpe_evaluating_2016, peng_topic_2020,hoernle_phantom_2022}.
Another interesting insight is that attitude change is often quantified via statistical-based techniques, such as in \citet{licorish_understanding_2014,tang_detecting_2020,ren_psychological_2022}.

With a similar aim, we explore existing combinations of behavior measures and change detection techniques in \Cref{fig:proxy_tech_heat}: count-based measures are often associated with visual analysis (such as in \citet{wang_anomaly_2015, jing_fine-grained_2020} and \citet{liu_monitoring_2021}) and statistical methods (such as in \citet{kibanov_mining_2017,liu_categorization_2020,petruzzellis_relation_2023}) for behavior change detection, while dictionary-based measures are generally examined in relation to statistical-based techniques (for instance in \citet{thelwall_sentiment_2011,stevens_desensitization_2021,foster_metoo_2022}).

The subject of behavior change is often a group: in most examples (such as \citet{jiang_topic_2011, tan_interpreting_2014, mcclellan_using_2017} and \citet{redondo_hybrid_2020}) this is linked to a group subject also in terms of behavior, but there are also examples of mismatches between behavior and behavior change subjects, for instance \citet{amelkin_distance_2019, kim_analysis_2021} and \citet{aljeri_big_2022}.
When the behavior change subject is an individual, the behavior of reference is always measured for the same individual, such as in \citet{liu_personalized_2010, makazhanov_predicting_2013} and \citet{uthirapathy_predicting_2022}.
Finally, a small number of works focus on studying behavior change at both individual and group levels: for instance, \citet{licorish_understanding_2014, kibanov_mining_2017, zygmunt_achieving_2020, dascalu_before_2021} and \citet{almerekhi_investigating_2022}.

The temporal level of behavior change is analyzed in the existing research equally in a discrete and gradual fashion, but most of the works focus on discrete change by detecting change points (such as \citet{jiang_topic_2011, nayak_exploring_2015, yeung_face_2020} and \citet{alattar_using_2021}) or by comparing a before and after event scenario (such as \citet{jiang_modeling_2015, kushner_bursts_2020, hashimoto_analyzing_2021} and \citet{chen_exploring_2022}).
Among the works exploring gradual behavior changes, there are examples of continuous analysis (for instance \citet{wang_anomaly_2015, franca_visualizing_2016, tahmasbi_go_2021} and \citet{lwin_evolution_2022}) and study of behavioral phases (for instance \citet{bui_temporal_2016, epure_modeling_2017, shi_temporal_2020} and \citet{choudrie_applying_2021}).
A minority of works study a combination of continuous and phase-defined gradual behavior changes as, for instance, in \citet{mavragani_infoveillance_2018, liu_categorization_2020, zhao_academic_2021} and \citet{alzamzami_monitoring_2021}, or a combination of discrete and gradual changes, such as in \citet{zygmunt_achieving_2020, zhao_academic_2021, flocco_analysis_2021} and \citet{dermy_dynamic_2022}.

\subsection{\rev{Theoretical Frameworks of Behavior Change}} \label{subsec:theories_behavior_change}

\subsubsection{\rev{Executive summary}}
\rev{
Less than 20\% of the existing literature explicitly incorporates concepts and theories from social or psychological sciences in the context of online behavior change. Among the most popular theoretical frameworks are emotional contagion (14.3\%) and social influence (10.7\%). 
}

\subsubsection{\rev{Detailed mapping}}
\rev{
Only a small percentage of the current research makes reference to social or psychological sciences concepts when dealing with behavior change: some works refer to concepts related to emotions, in particular in relation to emotion contagion \cite{shi_social_2020, naskar_emotion_2020, crocamo_surveilling_2021}, Plutchik's wheel of emotions \cite{basile_how_2021}, emotion awareness \cite{caliskan_how_2022}, or general emotion theory \cite{ngo_crowd_2016, ortega_modelling_2021, su_public_2021}.
Other popular social sciences concepts often mentioned are homophily \cite{strathern_identifying_2022}, social identity and polarization \cite{rogers_using_2021}, social change \cite{evers_covid-19_2021, balsamo_pursuit_2023} and influence \cite{zygmunt_achieving_2020, uthirapathy_predicting_2022}.
In terms of \rev{theories and theoretical frameworks} from sociology and psychology \rev{related to behavior change}, there is a reference to the affective events theory \cite{wang_anomaly_2015}, to Becker's marriage and Gale and Shapley's college admissions models \cite{dinh_computational_2022}, to the Five Doors Theory of behavior change \cite{harywanto_bertweet-based_2022}, to Gersick's punctuated equilibrium \cite{zhang_crowd_2017}, to professional learning community models \cite{scherz_whatsapp_2022}, to the regulatory focus theory \cite{abramova_collective_2022}, to the Kleinberg's hubs and authorities models \cite{achananuparp_who_2012}, to the moral foundations' theory \cite{mejova_authority_2023}, to the transtheoretical model of behavior change \cite{liu_assessing_2017} and to the goal-gradient hypothesis \cite{hoernle_phantom_2022}.
Finally, some works refer to other general theoretical concepts such as cognitive restructuring \cite{kushner_bursts_2020}, self-regulated learning \cite{zhang_measuring_2021}, affordances \cite{thelwall_sentiment_2011} and meaning-making vs. meaning-made processes \cite{foster_metoo_2022}.
}

\section{Discussion} \label{sec:discussion}
In this review, we critically examine the literature on observational studies of behavior change derived from online data, with the aim of shedding light on: \emph{Q1)} the types of \rev{behavior and} behavior change that research has focused on and the online platforms of reference; \emph{Q2)} the prevalent techniques for measuring \rev{behavior and} behavior change; and \emph{Q3)} the theoretical frameworks underpinning the empirical analysis of behavior change.

To answer these questions, we structure our discussion around three central elements common to all study designs: the online \emph{environment} of choice, the \emph{events} triggering behavior change, and the \emph{behavior} under analysis and whose change is measured.
Our findings indicate a significant concentration of research on a limited number of configurations of these elements.
The studies we reviewed disproportionately rely on Twitter or Reddit as data sources, a trend that raises concerns given the recent restrictions on their API usage~\cite{mashable2023apis}.
The COVID-19 outbreak is the event that has been investigated most often as a behavior change trigger.
The pandemic has served as a global-scale natural experiment, providing the ideal conditions for inferring causal relationships from observational data~\cite{conley2021past}, thereby driving much of the published research on behavior change from online data.
This surge in pandemic-related studies is attributable not only to the global significance of COVID-19 but also to the convenience of conducting straightforward \emph{pre-post} behavioral studies.
However, other equally important types of behavior change that originate within more articulated contexts, such as those related to climate action~\cite{rees2014climate}, remain understudied.
In terms of the behavior of focus, \rev{we adopted a broad an multidisciplinary perspective drawing from both behaviorism and cognitive psychology and considered both overt and covert behaviors.}
We found that emotions, sentiment, and topical interest expressed online are the most frequently analyzed behaviors, likely due to the wide availability of off-the-shelf tools for sentiment and topic mining~\cite{medhat2014sentiment,vayansky2020review}.
So far, less attention has been paid to more complex psychological and cultural concepts, including aspects of social pragmatics~\cite{gergen1985social,habermas2014pragmatics} (e.g., the process of gaining trust towards individuals or institutions) and political opinion formation, which constitute some of the core staples of Computational Social Science research~\cite{lazer2009computational}.
Furthermore, research aiming at detecting offline behavior change through online data is rarely observational and often relies on intervention-based experimental design.
The methods used to track behavior changes are typically simple numeric or visual assessments of temporal trends, often lacking even the support of basic statistical tests.
Last, in the realm of online behavior change research it is noticeable that the incorporation of a comprehensive theoretical framework remains somewhat deficient, thus potentially limiting the depth and precision of analytical insights. 
While certain studies draw upon concepts such as emotional contagion and social influence to explore behaviors and their transformations, it is important for future research to embrace a broader spectrum of psychological and social theories to provide a more exhaustive understanding of online behavioral dynamics.

The study by~\citet{valdez_social_2020} provides a compelling illustration of the prevailing research approach in this field. The authors examine the evolution of public discourse on Twitter during the first quarter of 2020, a period marked by the onset of the COVID-19 pandemic.
They compare the volume, topics, and sentiment of Twitter conversations before and after the pandemic outbreak.
The shift in topical focus is assessed through visual analysis of the results from the application of Latent Dirichlet Allocation (LDA), whereas the shifts in tweets volume and sentiment are assessed respectively via time series decomposition and Pruned Exact Linear Time algorithm for time series change points.
Similarly, \citet{tan_interpreting_2014} offers another influential contribution to the understanding of sentiment shifts on Twitter.
Although their work does not primarily focus on event-triggered behavior changes, they examine public sentiment towards specific entities, such as \emph{Obama} or \emph{Apple}.
They employ a dictionary-based tool to annotate tweets with sentiment labels and use the proportion of positive or negative tweets as an indicator of sentiment shifts.
They then apply LDA-based approaches to infer qualitative explanations of the observed shifts.
Our review has also revealed some less conventional, yet very promising approaches to quantifying online behavior change.
For example, \citet{ren_psychological_2022} use Glassdoor reviews as an alternative source of data to measure the psychological impact of the COVID-19 pandemic on people, particularly employees.
Using z-scores, they compare ratings and psycho-social indicators from dictionary-based analysis of the 2020 reviews with a baseline from 2017 to 2020.
An example of research that is extensively informed by sociological theories is the work by \citet{zhang_crowd_2017}, who explore group collaboration on Wikipedia.
Drawing on Gersick's punctuated equilibrium theory~\cite{gersick_revolutionary_1991}, they measure group dynamics in terms of activity level, centralization, conflict, experience level, machine-predicted quality, and sentiment.
By inspecting regression slopes across different periods of interest, they show that the collaborative behavior of editors of Wikipedia articles awarded with the Good Article (GA) status significantly changes just prior to the status nomination.
An example of non-traditional change detection techniques is given by \citet{uddin_impact_2014}.
By focusing on a student email network, they perform a score comparison of basic network measures over time, and they fit exponential random graph models to test the confidence significance of micro-structure parameters (e.g., edge and 2-star) as predictors in different time periods.

\spara{\rev{Limitations.}}
\rev{
This review adopts an multidisciplinary framework that incorporates behavioral, cognitive, and affective dimensions of behavior. While this broad perspective enables a richer understanding of online behavioral expression, it may sacrifice theoretical precision by favoring conceptual coverage over strict adherence to specific disciplinary definitions. This choice reflects the multidisciplinary nature of Computational Social Science but limits deep engagement with any single theoretical tradition.}

\rev{We focus exclusively on observational studies, which dominate empirical research in the field. While this allows us to map current measurement practices, it excludes experimental and interventional work that may offer stronger causal insights. Observational studies also face challenges in causal inference, particularly in complex online settings.}

\rev{Finally, our literature search is limited to the Scopus database. While Scopus provides extensive peer-reviewed coverage, it may omit relevant preprints, conference proceedings, or non-indexed journals. We view this trade-off as appropriate for establishing a stable foundation of published research, though future work could broaden the scope to capture more diverse and emerging contributions.}

\spara{Looking Forward.}
Our map of the state-of-the-art aims to identify gaps in the field that could inspire future research. We propose three key areas that we believe this field should address. 

First, we advocate for the expansion of behavior change studies across a broader spectrum of online platforms.
The sustainability of this research branch is threatened if it continues to rely solely on its conventional data sources.
Diversifying these sources will not only reduce dependence on individual data providers but also enhance the representativeness of computational research on human behavior by incorporating a variety of online environments that cater to diverse purposes and demographics.
Several alternative data sources are readily available, yet underexplored.
The decentralized social network Mastodon, which provides a rich and flexible API, offers a user experience akin to Twitter and is increasingly becoming a digital haven for users who became disenchanted with Twitter, following its acquisition by business tycoon Elon Musk~\cite{la2023get}.
The YouTube API~\cite{arthurs2018researching} offers a fertile ground for exploring the nexus between social movements and online discourse~\cite{abisheva2014watches}.
TikTok, which recently launched its first Research API, is a growing platform that particularly appeals to younger users and, unlike conventional social media, has a predominantly female user base~\cite{iqbal2021tiktok}.
Given its widespread popularity and diverse topics, TikTok is well-suited for studying a wide array of \rev{behavior and} behavior change types, from health habits to political participation~\cite{montag2021psychology,medina2020dancing}.
Public group interactions on instant messaging applications such as WhatsApp and Telegram also hold promise for studying alternative forms of activism and political participation, as they often serve as coordination tools for offline activities~\cite{garimella2018whatapp,gil2021whatsapp,urman2022they}.
However, it is important to note that most of these alternative services and APIs are proprietary and thus subject to any restrictions the platform owners may impose in the future.
Moreover, even when APIs are accessible, their restrictive terms of service could significantly hinder the sustainability of Open Science practices of transparency and reproducibility~\cite{davidson2023social}.
In this context, we urge social media platforms to acknowledge the crucial role they can play in advancing our understanding of behavior change in response to the collective challenges that necessitate widespread action and coordination.
We call on these platforms to adopt a more open research approach, offering enhanced and more transparent data access for research purposes.

Second, we encourage broadening the research perspectives in \rev{behavior and} behavior change studies and moving beyond the conventional focus on emotions and sentiment as the object of behavior.
Many pressing research questions in this field, which hold the potential for significant societal impact, remain way off the most beaten paths of research.
The analysis of large-scale, crowd-sourced data presents a promising avenue for understanding how to catalyze positive shifts in opinion, collective action, consensus, and cooperation---all of which are crucial for addressing pressing global challenges.
The lack of robust techniques for measuring these complex behavior shifts on a large scale partly explains the disproportionate emphasis that the existing literature has placed on simpler constructs such as sentiment.
However, recent years have seen remarkable progress in the development of novel methods for stance detection~\cite{kuccuk2020stance} and opinion mining~\cite{sun2017review,messaoudi2022opinion}.
Furthermore, the rapid evolution of Large Language Models (LLMs) such as OpenAI's GPT-4~\cite{openai2023gpt4} and Meta's LLaMA~\cite{touvron2023llama} has provided unprecedented opportunities for accurately capturing the nuanced types of human behavior through conversational language~\cite{ziems2023can}.
We urge future research to experiment with these models to broaden the spectrum of behavioral types \rev{and their shifts} to be studied systematically, and to eventually incorporate these models into their methodological toolkit.

Finally, in registering a significant disconnect between empirical research on behavior change and relevant theoretical frameworks, we recommend to enhance the collaboration between data science practitioners and social science experts.
In particular, we believe that the integration of theoretical frameworks in guiding data-driven studies will become increasingly vital as computer-assisted tools for behavior quantification gain more expressive power.

\spara{\rev{Considerations on Ethical Practices.}}
\rev{Working with individual data from online platforms raises important ethical issues, including informed consent for data collection, storage, and analysis—especially given the possibility of re-identification even when data is aggregated. 
Ethical approval for research involving identifiable data is essential, and such data should only be used for the approved purposes and accessed by authorized personnel. 
Additionally, potential harm and misuse of data, as well as risks of bias and discrimination arising from its use, must be carefully considered and acknowledged during analysis. 
Researchers should thoroughly assess these risks and include a clear ethical statement outlining their data handling practices in their publications.}

\section{Methods}\label{sec:methods}

\subsection{Literature Search Strategy} 
Our primary source of information was \emph{Scopus}, which we accessed on 7th February 2023.
\rev{We focused on Scopus due to its broad disciplinary coverage and reliable indexing of peer-reviewed literature in both social and computational sciences. While this may exclude some non-indexed work (e.g., preprints or conference proceedings), it ensures analytical depth and consistency in capturing trends within the published research landscape.}
To retrieve relevant literature, we used the guidelines of the Preferred Reporting Items for Systematic Reviews and Meta-Analyses (PRISMA) framework~\cite{page_prisma_2021} and followed a three-step search strategy:
\begin{enumerate}[leftmargin=*]
    \item We compiled a set of relevant keywords to compose a search query, and filtered search results with additional criteria to keep the corpus of retrieved results both focused and within a manageable size.
    \item We established eligibility criteria to direct the inclusion or exclusion process of the retrieved papers. Initial evaluation of these criteria was based on the title, abstract, and keywords of the papers. We then proceeded to assess the full text of the remaining articles against these criteria.
    \item We expanded the pool of potential papers with pertinent work cited by any of the ``literature review'' papers retrieved during our search. We also included more recent papers recommended by domain experts and published subsequent to our search process.
\end{enumerate}

\subsubsection{Query definition}
We defined the list of search keywords through an iterative process informed by domain experts. To keep a good level of recall without sacrificing precision, we matched our search keywords with only the title, abstract, and paper keywords. 
We compiled four sets of keywords: behavior-related, change-related, measurement-related, and online-related. 
\rev{The choice of behavior-related keywords reflects the broad an multidisciplinary perspective adopted by this review, including both overt and covert behaviors that are relevant in Computational Social Science and interactions in digital spaces.}
\rev{Rather than including terms like ``observational study'' directly in the query—which may be inconsistently used across papers—we opted to exclude works mentioning intervention-related terms. This ensures we capture a broader range of relevant observational studies, even when they are not explicitly labeled as such.}
The resulting query is the following:

\vspace{10pt}

\begin{mdframed}[backgroundcolor=gray!7,roundcorner=10pt]
\small
\begin{lstlisting}[breaklines]
(behavior* OR opinion* OR attitude* OR sentiment OR emotion* OR mood OR stance* OR belief) 
AND (change* OR changing OR shift* OR modification* OR alteration* OR variation* OR swing* OR drift*)
AND (measure* OR measuring OR operationalize OR operationalization* OR operationalizing OR quantify* OR quantification* OR detect* OR monitor* OR identify* OR identification* OR predict* OR estimate ORestimation* OR estimating OR assess* OR track*) 
AND (online OR social media OR twitter OR forum OR reddit OR youtube OR mastodon OR facebook OR whatsapp OR telegram OR blog*) 
AND NOT (intervention*)
\end{lstlisting}
\end{mdframed}

\subsubsection{Search filters} \label{subsub:filters}
To stike a good balance between recall and the number of retrieved papers to undergo the screening phase, we used the following search filters when submitting our query to \emph{Scopus}:
\begin{itemize}[leftmargin=*]
    \item Open access articles, to ensure accessibility of the sources.
    \item Articles published since the year 2000, to align with the time of the creation of the first blogs, given our review's focus on online platforms and social media.
    \item Articles within the realm of Computer Science, to enhance recall by retrieving articles that apply computational methods to data from online platforms. We also considered documents tagged with multiple disciplines, provided Computer Science was included.
    \item Document types were restricted to articles, conference papers, reviews, and conference reviews, to limit the inclusion of grey literature and maintain a manageable pool of candidate papers.
\end{itemize}

\subsection{Eligibility criteria}
We conducted a preliminary screening of the collected papers to identify and discard near-duplicates, retaining only the most recent publication in the case of multiple versions were found.
The selection of pertinent articles was conducted in two stages, first by inspecting the title, abstract, and keywords, and then by examining the full text of the article.
Given the varying levels of information detail across these two stages, we established two distinct sets of eligibility criteria designed to complement the filtering achieved by the search query.
These criteria are outlined in \Cref{tab:criteria_measure}. 
Throughout both stages, the lead author screened all articles, with a subset of documents independently reviewed by the last author.
During the first screening phase (based on title, abstract, and keywords), 30\% of the articles \rev{were independently reviewed} by the last author. 
The percentage agreement between the two authors is almost 95\%, and Cohen's Kappa is 0.88.
Similar results are found in the second screening phase, in which 15\% of full-text papers undergo a double independent screening. 
Any disagreements were resolved through discussion.

\begin{table}
    \centering
    \caption{Eligibility Criteria \rev{for works to be included in the study.}}
    \label{tab:criteria_measure}
    \setlength{\leftmargini}{0.4cm}
    \begin{tabularx}{\textwidth}{L{0.1\textwidth} L{0.2\textwidth} L{0.65\textwidth}}
        \toprule
        \textbf{Screening phase} & \textbf{Inclusion} & \textbf{Exclusion} \\
        \midrule
        \textbf{Title, abstract, keywords} & 
        The study is conducted, at least partially, on online platforms & 
        \begin{itemize}[nosep,after=\vspace*{-\baselineskip}] \vspace*{-.6\baselineskip}
            \item Purely theoretical work
            \item Focus on the development of apps/interfaces
            \item Type of behavior not of interest (e.g., non-human behavior, stock market)
            \item Behavior retrieval does not match the survey's scope (e.g., active measure, IoTs)
            \item Task of reference is not behavior change measurement (e.g., recommendation systems study, pure topic modeling)
        \end{itemize} \\ 
        \midrule
        \textbf{Full-text} & The study reports a clear description of at least one measurement technique for behavior change & 
        \begin{itemize}[nosep,after=\vspace*{-\baselineskip}] \vspace*{-.6\baselineskip}
            \item The study was to be excluded during the first screening phase but exclusion criteria fulfillment was not clear prior to full-text screening
            \item MSc/BSc thesis or Ph.D. dissertations
            \item Focus is on comparing behavior between groups and not on change
            \item Focus is on algorithms for feature extraction and not on behavior change
            \item Qualitative analysis of data without main focus on behavior change
        \end{itemize} \\
        \bottomrule
    \end{tabularx}
\end{table}

\subsection{Retrieved Papers}

\begin{figure}[ht!]
  \centering
  \includegraphics[width=1\linewidth]{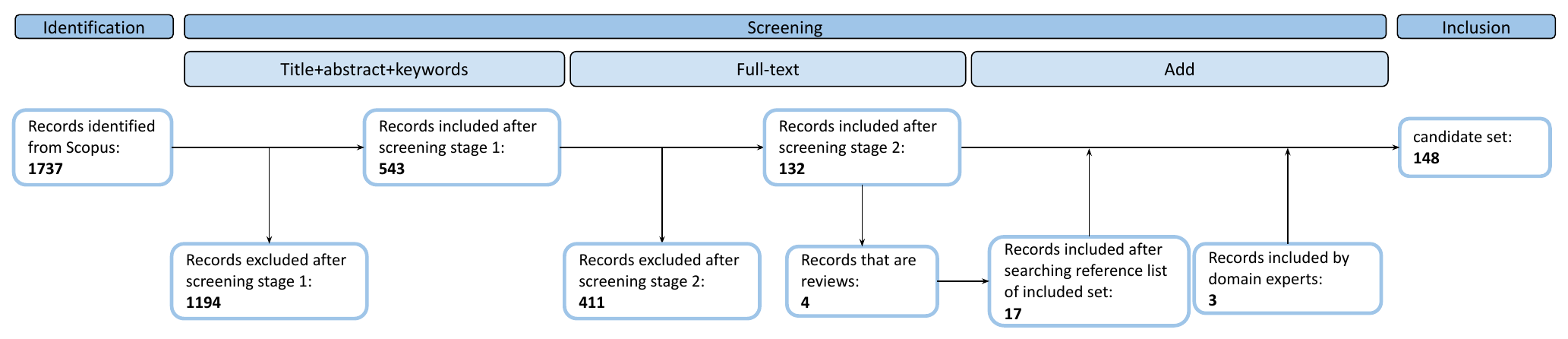}
  \caption{PRISMA search strategy flowchart.}
  \label{fig:retrieved_numbers}
\end{figure}

\Cref{fig:retrieved_numbers} illustrates the progression of the literature search process, detailing the number of papers retained at each stage.
Initially, the search query and initial filtering process (described in \S\ref{subsub:filters}) yielded 1737 papers.
Following the first round of screening, 543 papers were retained.
These were then subjected to a second screening stage, resulting in a shortlist of 132 papers, including four review articles.
The final set of candidate papers, totalling 148, was achieved by incorporating relevant references from the initial set and recommendations from subject matter experts. 
\Cref{fig:years_numbers} provides a year-wise breakdown of the papers retrieved, highlighting an increasing interest in the operationalization of online behavior change staring from 2020.

\section{Related Work} \label{sec:related}

The wealth of data from online platforms has opened up numerous avenues for observational studies on human behavior.
However, most existing literature reviews on methods for measuring behavior cover primarily intervention-based experiments and largely focus on health behavior change. 
\citet{maher_are_2014} reviewed the existing literature on the effectiveness of health behavior interventions delivered through online social media.
In that review, behavior change is consistently quantified by means of effect size-based analyses on the controlled trials, and some of the retrieved studies are linked to social theories of behavior change. 
\citet{lovato_impact_2011} conducted a similar analysis, albeit with a focus on traditional mass media.
They aimed to assess the impact of mass media advertising on the smoking behaviors of young adults through a review of longitudinal studies.
Most of the studies covered in the review measure change with a binary assessment of whether the subjects were still smoking during the follow-up stage.
\citet{li_vaccine_2022} examined the issue of social media interventions to counter vaccine hesitancy.
The studies they reviewed used both active and passive measurements of behavior change, including self-reported surveys, interviews, and data extraction from medical records.
Still in the realm of health behavior interventions, \citet{ridout_use_2018} focus on youth mental health and find a small number of quantitative studies compared to the set of qualitative ones.
Mental health is also taken into account by \citet{skaik_using_2020}, reviewing works focused on using machine learning models for the prediction of mental disorders or on the study of mental health evolution.
These models often rely on ground truth data from human annotation, questionnaires, or self-disclosure~\cite{wongkoblap_researching_2017, chancellor_methods_2020}, and are trained on features based on textual, multimedia, and behavioral user characteristics~\cite{dhelim_detecting_2023}. 
While the surveys mentioned here represent only a small selection of reviews on health behavior change analyses, they provide a comprehensive overview of change quantification methods for such studies. However, as our survey focuses on behavior change in a broader sense, we do not delve further into the existing health-focused literature.

Our work partially intersects with the sub-fields of opinion change and dynamics.
A review by~\citet{cortis_over_2021} concentrates on identifying opinion dimensions from social media platforms.
The authors categorize existing opinion dimensions and touch upon sentiment change, without delving into the methods of measuring such phenomena.
\citet{grabisch_survey_2020} examine models of opinion dynamics from a theoretical perspective, often incorporating features related to opinion change; however, they overlook the experimental application of such models on online platforms.
\citet{veenstra_networkbehavior_2013} offer a compelling review that covers the dynamics of both behavior and opinion, emphasizing the impact of social networks and the role of influence on individual behaviors. 
The effort is, however, limited to models of behavior dynamics and does not involve online data.
Finally, \citet{li_survey_2017} provide a broad overview of existing models of information diffusion applied to online social networks.
Even though opinion and behavior dynamics studies are related to the topic of behavior change, the purpose of our survey is to focus on the quantification of behavioral shifts without delving into the modeling of social network dynamics.

In summary, existing literature reviews predominantly focus on health-related behaviors, intervention-based studies, and theoretical models, with a noticeable absence of reviews on methods for measuring various types of behavior change in observational studies on social media and other online platforms. Our work proposes to fill this gap by providing an overview of current trends, developing a classification of behavior change measures, and identifying unexplored areas to stimulate future research.

\newpage
\bibliographystyle{ACM-Reference-Format}
\bibliography{references,ref_luca}

\end{document}